\documentclass[twocolumn]{aastex61}
\shorttitle{HST/STIS Observations of GJ436\MakeLowercase{b}}
\shortauthors{Lothringer et al.}

\usepackage{amsmath}
\usepackage{xspace}
\usepackage[normalem]{ulem}

\def\mearth{{\rm\,M_\oplus}}

\def\rearth{{\rm\,R_\oplus}}

\newcommand{\microns}{$\mu$m}

\newcommand{\spitzer}{{\it Spitzer}}
\newcommand{\fhaze}{{$f_{haze}$}}

\begin{document}
\title{An HST/STIS Optical Transmission Spectrum of Warm Neptune GJ 436\MakeLowercase{b}}
\author[0000-0003-3667-8633]{Joshua D. Lothringer}
\affiliation{Lunar and Planetary Laboratory, University of Arizona, Tucson, AZ, USA}
\author[0000-0001-5578-1498]{Bj{\"o}rn Benneke}
\affiliation{Division of Geological \& Planetary Sciences, California Institute of Technology, Pasadena, CA, USA}
\affiliation{Universit\'e de Montr\'eal, Montr\'eal, QC, Canada}
\author[0000-0002-1835-1891]{Ian J. M. Crossfield}
\affiliation{Department of Physics, Massachusetts Institute of Technology, Cambridge, MA, USA}
\affiliation{Astronomy and Astrophysics Department, UC Santa Cruz, CA, USA}
\author[0000-0003-4155-8513]{Gregory W. Henry}
\affiliation{Center of Excellence in Information Systems, Tennessee State University, Nashville, TN, USA}
\author[0000-0002-4404-0456]{Caroline Morley}
\affiliation{Harvard University, Cambridge, MA, USA}
\author[0000-0003-2313-467X]{Diana Dragomir}
\affiliation{Kavli Institute for Astrophysics and Space Research, Massachusetts Institute of Technology, Cambridge, MA, USA}
\author[0000-0002-7129-3002]{Travis Barman}
\affiliation{Lunar and Planetary Laboratory, University of Arizona, Tucson, AZ, USA}
\author[0000-0002-0822-3095]{Heather Knutson}
\affiliation{Division of Geological \& Planetary Sciences, California Institute of Technology, Pasadena, CA, USA}
\author[0000-0002-1337-9051]{Eliza Kempton}
\affiliation{Department of Physics, Grinnell College, Grinnell, IA, USA}
\author[0000-0002-9843-4354]{Jonathan Fortney}
\affiliation{Astronomy and Astrophysics Department, UC Santa Cruz, CA, USA}
\author[0000-0001-9165-9799]{Peter McCullough}
\affiliation{Space Telescope Science Institute, Baltimore, MD, USA}
\author[0000-0001-8638-0320]{Andrew W. Howard}
\affiliation{Astronomy Department, California Institute of Technology, Pasadena, CA, USA}
\vspace{0.5\baselineskip}
\date{\today}
\email{jlothrin@lpl.arizona.edu}

\begin{abstract}
	%Full abstract
GJ 436b is a prime target for understanding warm Neptune exoplanet atmospheres and a target for multiple JWST GTO programs. Here, we report the first space-based optical transmission spectrum of the planet using two HST/STIS transit observations from 0.53-1.03 $\mu$m. We find no evidence for alkali absorption features, nor evidence of a scattering slope longward of 0.53 $\mu$m. The spectrum is indicative of moderate to high metallicity ($\sim 100-1000 \times$ solar) while moderate metallicity scenarios ($\sim 100 \times$ solar) require aerosol opacity. The optical spectrum also rules out some highly scattering haze models. We find an increase in transit depth around 0.8 \microns \xspace  in the transmission spectra of 3 different sub-Jovian exoplanets (GJ 436b, HAT-P-26b, and GJ 1214b). While most of the data come from STIS, data from three other instruments may indicate this is not an instrumental effect. Only the transit spectrum of GJ 1214b is well fit by a model with stellar plages on the photosphere of the host star. Our photometric monitoring of the host star reveals a stellar rotation rate of 44.1 days and an activity cycle of 7.4 years. Intriguingly, GJ 436 does not become redder as it gets dimmer, which is expected if star spots were dominating the variability. These insights into the nature of the GJ 436 system help refine our expectations for future observations in the era of JWST, whose higher precision and broader wavelength coverage will shed light on the composition and structure of GJ 436b's atmosphere.

\end{abstract}


\section{Introduction}
Found close to their host star with short periods and large radii, hot Jupiters are among the easiest targets for characterization through transit spectroscopy and have been studied in increasing detail in recent years. However, hot Jupiters represent only a small fraction of the greater exoplanet population \citep{dressing:2013,fressin:2013,foreman-mackey:2014,mulders:2015}. More prevalent smaller and less massive planets are now being studied using observing practices and data analysis techniques developed from the study of hot Jupiters. Distinct from hot Jupiters are objects that more closely resemble the Solar System's ice giants in mass and radius. However, unlike our Solar System, these exoplanets can be found at short orbital periods with relatively high temperatures. This class of exoplanet, sometimes called warm Neptunes, occupy orbital periods on the order of days, orbital distances of less than 0.1 AU, and temperatures of $\sim 500-1000$ Kelvin.

An important question to ask about warm Neptune exoplanets is to what degree they resemble our own Solar System's ice giants. Planet formation, orbital evolution, atmospheric evolution (especially through atmospheric escape), and past and present stellar irradiation may all play significant roles in shaping their atmospheres as we see them today. Thus characterizing the atmosphere of these planets offers a path to test and improve models of planet formation and evolution. To that end, much effort has gone into modeling and observing the atmospheres of sub-Jovian planets \citep[e.g.,][]{kreidberg:2014,fraine:2014,knutson:2014a,knutson:2014b,benneke:2017,moses:2013,benneke:2012,benneke:2013,miller-ricci:2009,miller-ricci:2010}. 

To date, spectroscopic observations have been made of only about a dozen sub-Jovian systems. An even smaller number have measured water absorption identified in their near infrared transmission spectrum, all of which require either a cloudy or high metallicity atmosphere to explain the observations \citep{fraine:2014,stevenson:2015,wakeford:2017b}. Other sub-Jovian exoplanets have observed spectra devoid of water absorption, again either due to obscuring clouds or high metallicity \citep{knutson:2014a,knutson:2014b,kreidberg:2014}. \cite{crossfield:2017} recently suggested a correlation between the depth of the water absorption feature and either the planet's equilibrium temperature or H/He mass fraction. 

\subsection{High Mean Molecular Mass Atmospheres versus Clouds and Hazes}
Aerosols may form in the atmosphere of sub-Jovian exoplanets as clouds of condensates or as hazes of photochemical products (e.g., hydrocarbons). Aerosols in the atmospheres of sub-Jovian exoplanets have long plagued transit observations due to their ability to mute potential absorption features; the long path length through the atmospheres from the geometry of transit observations amplifies this problem significantly \citep{fortney:2005}. This issue can be especially troublesome for small exoplanets, where small scale heights and planet-to-star radius ratios can lead to intrinsic variations of transit depth with wavelength that are already small (i.e., on the order of the systematic noise), even in an aerosol-free scenario. In some exoplanet studies, clouds are defined as a gray (wavelength-independent) opacity source, while hazes are defined as a scattering opacity source that can induce slopes in the spectrum. For this work, we define clouds and hazes based on their formation and physical properties (i.e., clouds are condensed aerosols, while hazes are aerosols formed through photochemical processes), not on their effect on the planet's spectrum. In the models we use to interpret our data, whether an aerosol acts as a gray or scattering opacity depends primarily on the particle size rather than its classification as a cloud or haze. 

High metallicity atmospheres can affect the planet's spectrum in similar ways to aerosols. High metallicity results in a high mean molecular mass atmosphere, causing a reduced atmospheric scale-height. This reduced scale-height results in a smaller signal in the transmission spectrum, serving to mute spectral features. High metallicity can be a natural outcome of the formation for low-mass planets through core-accretion \citep{fortney:2013,thorngren:2016,venturini:2016}. Though distinguishing between high-metallicity and aerosol-rich atmospheres is difficult and requires high signal-to-noise data, \cite{benneke:2013} provided a framework to do so by measuring line wing steepness and the relative absorption depth of different spectral features. Additionally, the transmission spectrum of high metallicity atmospheres and atmospheres with aerosols begin to diverge at both short and long wavelengths, providing an opportunity to break this degeneracy (see Section \ref{jwst}). 

Observations at optical wavelengths provide a path forward in studying clouds, hazes, and high metallicity exoplanet atmospheres in transit. As mentioned above, scattering by small aerosol particles can dominate the optical spectrum by producing a slope toward larger transit depth at shorter wavelengths. Cloud-free atmospheres, on the other hand, have transmission spectra that reach minimum transit depths near 0.5 \microns \xspace before Rayleigh scattering dominates shortward of 0.5 \microns \xspace and molecular opacities dominate longward of 0.9 \microns.  Meanwhile, atmospheres with large aerosol particles will remain flat at optical wavelengths. Thus the optical spectrum provides a unique way to characterize opacity sources even in an otherwise featureless spectrum. Additionally, absorption from atomic Na and K can shape much of an exoplanet's optical transmission spectrum \citep[e.g.,][]{seager:2000,nikolov:2014,sing:2015}. Characterization of these features can provide a measurement of the atmosphere at lower pressures than those probed by the infrared spectrum \citep{sing:2008,vidalmajar:2011,vidalmajar:2011corr,heng:2015,wyttenbach:2015}. The absence of Na and K in the optical spectrum of an exoplanet may indicate the condensation of these elements into clouds (e.g.,~KCl and Na$_2$S \citep{morley:2013}).

Caution is necessary when interpreting slopes at optical wavelengths in transmission spectra as occulted and unocculted star spots on the photosphere of the host star can produce slopes in the optical transmission spectrum \citep[][see Sections \ref{stellar_activity_effects}]{berta:2011,oshagh:2013,mccullough:2014,oshagh:2014,rackham:2017}. Photometric monitoring can reveal changes in the star spot or plage filling factor as well as the overall activity cycle. Brightness modulations from stellar rotation, star spot variability, and activity cycles all need to be accounted for in secondary eclipse and phase curve observations (see Section \ref{PHOTO}). Spectroscopic monitoring provides information on the host star's absolute activity level via activity indicators like the Ca II H and K lines.

\subsection{GJ 436b: The First Warm Neptune}

GJ 436b is a 21.4 $\mearth$ (1.25 M$_{Neptune}$, 0.0673 M$_{Jupiter}$) warm Neptune with a radius of 4.2 $\rearth$ (1.1 R$_{Neptune}$, 0.37 R$_{Jupiter}$) \citep{trifonov:2017,turner:2016}. Discovered by radial velocity by \cite{butler:2004} as the first Neptune-mass exoplanet, it was subsequently found to transit by \cite{gillon:2007b}. Transit spectra were first obtained by \citep{pont:2009} using NICMOS on the Hubble Space Telescope (HST), placing upper limits of a few parts per 10,000 on the potential water absorption feature at 1.4 $\mu$m. HST/WFC3 transit spectra also revealed a featureless spectrum, ruling out a cloud-free, hydrogen-dominated atmosphere \citep{knutson:2014a}. Using \spitzer \xspace transit measurements, \cite{beaulieu:2011} claimed to detect CH$_4$ and found no evidence of CO or CO$_2$ due to large absorption measured in the 3.6 and 8 $\mu$m IRAC bands. This was refuted by \cite{knutson:2011} who hypothesized that stellar activity caused the spectrum to vary not only in wavelength, but also with time. Reanalysis of these data by \cite{lanotte:2014} and \cite{morello:2015} using new detrending techniques found that the transmission spectrum of GJ 436b was constant with wavelength and did not vary between epochs. 

GJ 436b's dayside spectrum has been observed through secondary eclipse measurements with \spitzer. GJ 436b's equilibrium temperature of 700-800 K would imply CH$_4$ is the most abundant carbon-bearing molecule in chemical equilibrium; however, measurements from \cite{stevenson:2010} suggested that CO was in high abundance, rather than CH$_4$. Additional studies have supported the assertion that GJ 436b is enhanced in CO and CO$_2$ and deficient in CH$_4$ \citep{madhusudhan:2011c,agundez:2012,lanotte:2014}. This could potentially be explained by disequilibrium processes like vertical mixing and tidal heating. Photochemistry is likely not the cause of this CH$_4$ deficiency; \cite{line:2011} estimate that the rate at which CH$_4$ should be destroyed was much less than could explain the observations.

A more recent analysis including both self-consistent modeling and retrievals of the emission and transit spectra placed a 3-$\sigma$ lower limit on the metallicity at 106$\times$ solar \citep{morley:2016}. A cloud-free atmosphere is still possible if metallicity is on the order of 1,000$\times$ solar; for lower metallicities, clouds are needed to help mute spectral features to match observations. Additionally, \cite{morley:2016} found that models with disequilibrium chemistry through quenching of CH$_4$, CO, and CO$_2$ with enhanced internal heating, presumably from tides, best matched the data. These results tend to agree with previous modeling of Neptune-sized planets from \cite{moses:2013}, who found CO enrichment and CH$_4$ depletion to be a natural consequence of high metallicity. In this present work, we present new observations at optical wavelengths in order to test the conclusions of \cite{morley:2016}.

At present time, GJ 436b is a candidate target for the NIRISS, NIRCam, and MIRI JWST Guaranteed Time Observation (GTO) programs. These programs will observe multiple secondary eclipses of the planet from 0.7-11 \microns, providing an unprecedented look at the atmospheric composition and structure of GJ 436b's dayside \citep[see][]{greene:2016}. We discuss how complementary transit observations will help distinguish cloudy and high-metallicity scenarios and determine conditions at the planet's terminator.

Here, we present the first space-based measurements of the optical transit spectrum of GJ 436b, interpret the full optical-to-IR spectrum, and constrain the rotation period and activity cycle of the host star. We organize the paper as follows: In Section 2 we explain the observations and data reduction procedures used. Section 3 describes our light curve fitting techniques. In Section 4, we show our results, including a look at the effects of stellar variability and the use of different orbital solutions. In Section 5, we compare our results to other sub-Jovian exoplanets and describe a common trend found in their optical transmission spectrum. We close with a discussion of expectations for future JWST transmission spectroscopy of GJ 436b.

\section{Observations and Data Reduction}

%  \begin{deluxetable}{ll} \label{table:dates}
% 	%\tablecolumns{2}
% 	\tablecaption{Observations of GJ 436b}
% 	\tablehead{\colhead{Visit Date} & \colhead{Instrument}}
% 	\startdata   
% 	26 October 2012 &  WFC3/G141 \\
% 	29 November 2012   &  WFC3/G141\\
% 	10 December 2012   &  WFC3/G141\\
% 	2 January 2013   &  WFC3/G141\\
% 	10 June 2015       &  STIS/G750L\\
% 	14 June 2015 &  STIS/G750L \\
% 	\enddata
% 	\label{tab1}
% \end{deluxetable}

\subsection{STIS Observations}

Two transits of GJ 436b were observed on 10 June 2015 UT and 14 June 2016 UT using the Space Telescope Imaging Spectrograph (STIS) on the Hubble Space Telescope (HST) with the G750L grism ($0.53-1.03 ~ \mu m$) as part of GO 13665 (PI Benneke). The 52x2 arsec$^2$ slit was used to minimize slit losses, while a 128 pixel subarray mode reduced readout overhead. Each transit consisted of 4 total HST orbits. The first orbit of a visit often exhibits strong systematic variations, inconsistent with the systematics in subsequent orbits. For this reason, after confirmation of this phenomenon, the first orbit was not included in the analysis. The second and fourth orbits are used to characterize the stellar baseline flux, while the transit occurs during the third orbit. Exposure times were 100 seconds resulting in 20 frames per HST orbit. The first frame of each orbit consistently shows anomalous flux, so this frame is not included in the light curve fitting. Both visits were scheduled with the same orbital phasing, with the in-transit HST orbit covering ingress through transit center to about 4 or 5 frames after transit center. 

A STIS pipeline was built for the program using existing CALSTIS routines in addition to custom procedures for cosmic ray identification. This pipeline was tested and validated using previous STIS data sets and is described below.

\subsubsection{Cosmic Ray Identification and Removal}

The long exposure times required for GJ 436 (I = 8.3) meant that most frames had multiple cosmic ray hits. It has also previously been found that the default CALSTIS routines for cleaning cosmic ray hits were inadequate for our purposes \citep{nikolov:2014}. For these reasons, a custom cosmic ray identification and removal procedure was developed, drawing on the technique described by \cite{nikolov:2014}. 
For each frame to be cleaned, four difference images were created between the frame and the four frames nearest in time. Each difference image will subtract out the stellar flux, leaving only the cosmic rays from one frame being positive values and the cosmic rays from the other frame being negative values. A median difference frame was then created from the 4 difference frames. The median difference frame consists only of cosmic ray hits. Next, each pixel's flux was compared to the standard deviation of its column. If the pixel value was greater than 4 times this standard deviation, it was flagged as a cosmic ray hit. As a second method, a window of 20x20 pixels was then placed on each pixel of the median difference frame and the median for that window was calculated. If the center pixel's value was greater than 4 standard deviations from this median window value, the center pixel was flagged as being contaminated by a cosmic ray strike. 

Once all pixels were analyzed in this manner, the pixel values were replaced by the corresponding pixel value in the median difference frame (i.e., the median for the four nearest frames). %Nikolov does this differently
Pixels identified as `bad' according to CALSTIS were replaced in this same manner. After being extracted from the 2 dimensional frames (see below), the 1D spectra were checked for cosmic rays that were missed by the 2D procedure. This is especially important for cosmic rays that occur near the bright spectral trace. The 1D stellar spectrum was compared to the two nearest spectra and two difference frames were created. If the 1D spectrum exceeded the average of these difference frames at any point by 10 standard deviations, that pixel was flagged as being contaminated by a cosmic ray strike. The large standard deviation cut-off serves to ensure that we are not erasing statistical noise. The value of the contaminated pixel is then replaced by the mean of the 2 nearest values.

\subsubsection{Spectral Extraction}

The data were dark subtracted, bias corrected, and flat fielded using the appropriate CALSTIS routines. The G750L grism on STIS has an obvious fringing effect longwards of 0.7 $\mu$m. To account for this, a fringe flat was taken at the end of each HST visit (i.e., one for each transit). This fringe flat was then used to divide out the fringe pattern using the CALSTIS defringing procedure \citep{goudfrooij:1998}. We found that defringing did not have a significant effect on the resulting planetary spectrum. 

Measurements of the spectral trace slope, the shift of the spectrum in the spatial direction and the shift of the spectrum in the spectral direction were saved to be later used as covariates in the light curve analysis for systematic detrending. The spectral trace slope and spatial shift were found by fitting a Gaussian profile to each column and then fitting a line through the peaks of all the profiles. The shift of the spectrum in the dispersion direction was measured by taking the wavelength value at the center column after image rectification in CALSTIS.
	
An extraction aperture size of 13 pixels was found to minimize the scatter in the residuals to the fit. Other extraction aperture sizes lead to equivalent results. Both the CALSTIS X1D routine and IRAF/APALL were used to extract the spectrum, each producing consistent results.
 
\subsection{Photometric Monitoring: A Precise Rotation Period} \label{PHOTO}
\begin{figure*}
	\begin{center}
		\includegraphics[width=.95\textwidth]{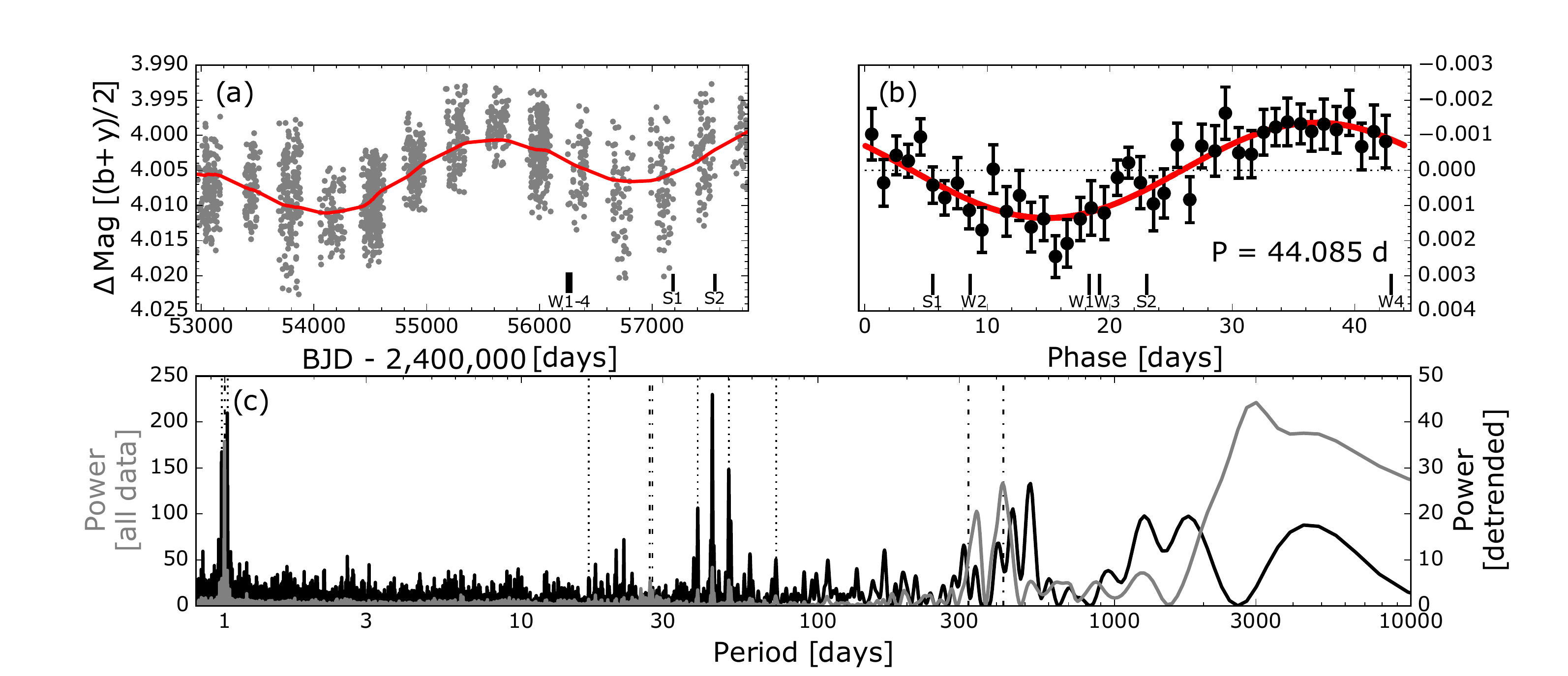}
	\end{center}
	\caption{Photometric monitoring of GJ~436.  Panel (a) shows the 14-year APT data set, along with a $\sim$7.4~yr model that likely indicates a stellar activity cycle. Panel (b) shows the photometry after removing the long-term trend, folding on the 44.1~d stellar rotation period, and binned to a one-day cadence.  The times of the STIS and WFC3 visits are indicated by ``S" and ``W", respectively. The periodograms of the raw (gray) and detrended (black) photometry are shown in panel (c), with vertical lines noting the daily, monthly, and yearly aliases of the rotation (dotted) and activity (dot-dashed) periods.  \label{fig:phot}}
\end{figure*}

\begin{figure}
	\begin{center}
		\includegraphics[width=.48\textwidth]{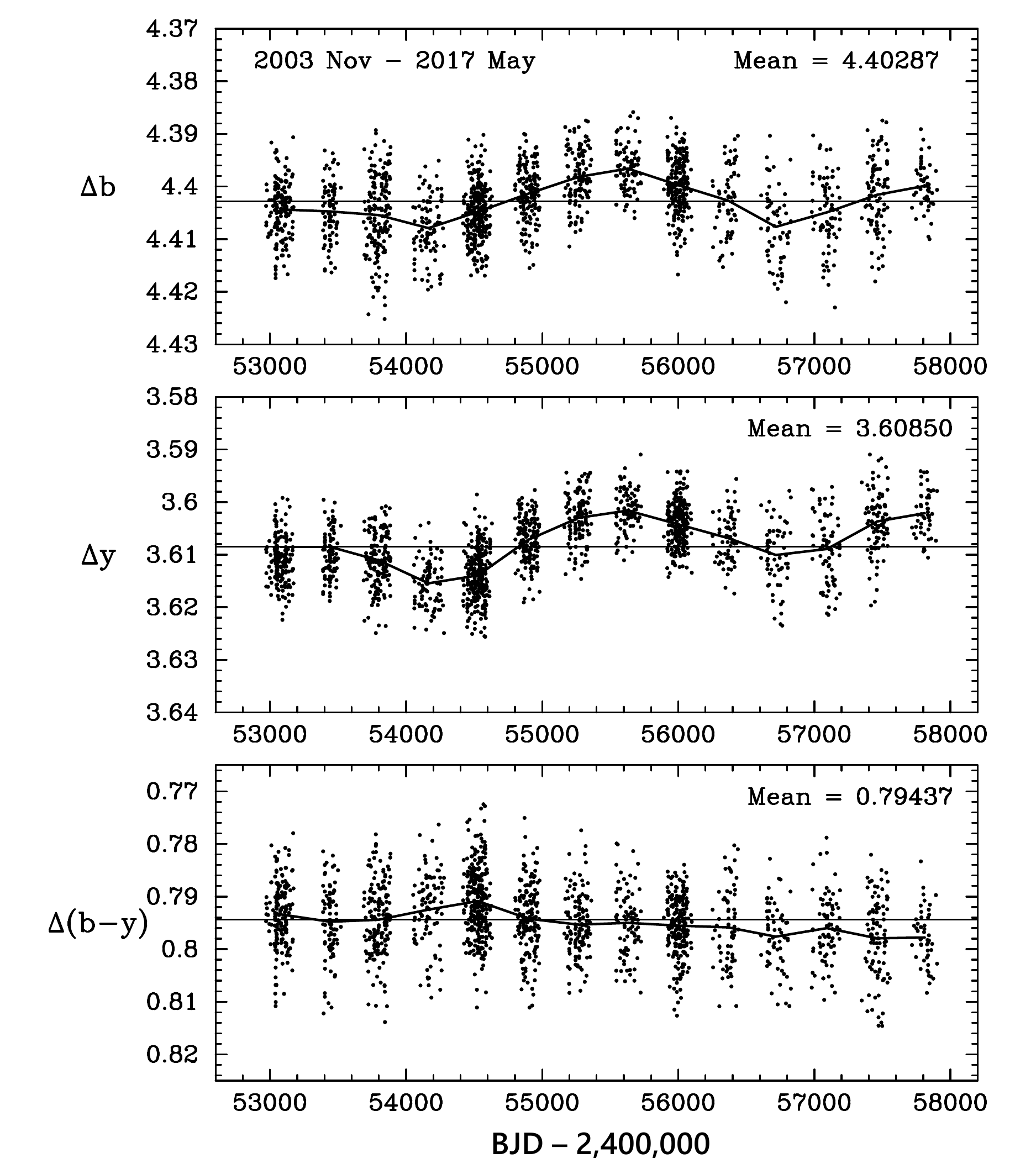}
	\end{center}
	\vspace{-20pt}
	\caption{Differential b (top), y (middle), and b-y (bottom) magnitudes for GJ~436 over 14 years computed with respect to the mean brightness of two comparison stars. GJ 436 does not exhibit the expected color variation for star spot dominated variability (see text). The standard deviation of the mean of the $\Delta$ (b-y) color in each observing seasons is around 0.001. \label{fig:color}}
\end{figure}

GJ~436 (M2.5V) has been monitored in Str\"omgren b and y filters for the past fourteen years using the Tennessee State University's T12 0.8 m Automatic Photoelectric Telescope (APT) at Fairborn Observatory in southern Arizona \citep{henry:1999,eaton:2003,henry:2008} in order to better characterize the star and how star spots may affect our transit observations. In these observations we nod the telescope between GJ 436 and comparison stars of comparable or greater brightness; we use these stars to remove extinction and seeing effects. Of the 3 comparison stars analyzed, HD 102555 (hereafter C1) and HD 103676 (C2) were the most constant, with very little variation between them over the 14 years. The third star, HD 99518, showed a gradual brightening throughout the 14 year observations. Altogether we have obtained 1735 measurements, which we present in Table \ref{photometrydata} and Figure \ref{fig:phot} and \ref{fig:color}. 

\begin{deluxetable*}{ccccc}
	\tabletypesize{\small}
	%	\tablenum{1}
	\tablewidth{0pt}
	\tablecaption{STR\"OMGREN PHOTOMETRIC OBSERVATIONS OF GJ 436b}
	\tablehead{
		\colhead{Date} & \colhead{(P$-$C1C2\tablenotemark{a})$_{b}$} & \colhead{(P$-$C1C2)$_{y}$} & \colhead{(P$-$C1C2)$_({b-y)}$} & \colhead{(P$-$C1C2)$_{(b+y)/2)}$} \\
		\colhead{(BJD$_{TDB}$ $-$ 2,400,000)} & \colhead{(mag)} & \colhead{(mag)} & \colhead{(mag)} & \colhead{(mag)}
	}
	\startdata
	52970.9856 & 4.40370 & 3.60900 & 0.79470 & 4.00635 \\
	52971.9748 & 4.40740 & 3.61605 & 0.79135 & 4.01175 \\
	52972.9752 & 4.40685 & 3.61350 & 0.79335 & 4.01020 \\
	52973.9700 & 4.39960 & 3.60950 & 0.79010 & 4.00455 \\
	52977.9596 & 4.40075 & 3.60855 & 0.79220 & 4.00470 \\
	52984.9665 & 4.40960 & 3.61610 & 0.79350 & 4.01285 \\
	\enddata
	\tablecomments{Table 1 is presented in its entirety in the electronic 
		edition of the Astronomical Journal.  A portion is shown here for 
		guidance regarding its form and content.}
	\tablenotetext{a}{C1C2 denotes that the differential magnitudes
		are computed with respect to the mean brightness of comparison
		stars C1 and C2}
	\label{photometrydata}
\end{deluxetable*}

The photometry of GJ~436 in Figure \ref{fig:phot} shows an obvious long-term variation with a period of roughly 7.4~yr and a peak-to-peak amplitude of 10~mmag.  This coherent signal is consistent in period and  amplitude with observations of many low-mass stars that are interpreted as stellar activity cycles \citep{mascareno:2016}. We therefore conclude that we have measured GJ~436's stellar activity cycle for the first time. This activity cycle is consistent with the finding that early M-type stars have magnetic cycles that are on average 6.0 $\pm$ $2.9$ years and mid-M-type stars average 7.1 $\pm$ $2.7$ years \citep{mascareno:2016}. %This activity cycle is also very similar to that of Proxima Centari (M5.5Ve), which was recently measured to be 7 years \citep{wargelin:2017}.  

After removing this signal and a linear trend, a periodogram analysis reveals a strong signal with a period of 44.1 $\pm$ 0.2~d and peak-to-peak amplitude of 3~mmag. To estimate the uncertainty on the rotation period, we split the data into 3-year blocks and computed a Lomb-Scargle periodogram of each data subset.  The 44~d signal is not detectable in each individual season's photometry, but a strong peak is visible in the periodogram for each of these five contiguous sub-blocks.  We measured the location of the peak of each of these five periodograms and find the mean and standard deviation of the mean to be $44.1 \pm 0.2$~d, which we interpret as GJ~436b's rotation period and the uncertainty on that parameter.

The rotational period we measure for GJ 436 is similar to periods observed in many low-mass stars \citep{irwin:2011,mascareno:2016}. This measurement is also consistent with an early rotation period derived from spectroscopic indicators \citep{demory:2007}, but inconsistent with the 57~d period derived using a one-year segment of our data \citep{knutson:2011}.  When using those same data we also see a 57~d periodogram peak, but that season is the only one of our fourteen whose periodogram shows a peak at that period.  We conclude that  GJ~436's rotation period is indeed 44.1$\pm$ 0.2~d. This rotation rate is consistent with the interpretation that GJ 436 is not an active star and is at least a few Gyr old \citep{maness:2007,kiraga:2007,saffe:2005,jenkins:2009,sanz-forcada:2010}. With respect to the relationship between rotational period and activity cycle, GJ 436 fits in with other M-stars \citep{mascareno:2016}.

Figure \ref{fig:color} shows the same photometric observations, but for the individual b and y filters, showing that as GJ 436 first gets dimmer, the star becomes bluer. Most chromospherically active stars get redder as the star gets fainter, implying that dark spots dominate the variability \citep[e.g.,][]{innis:1997}; however, another well-known chromospherically active RS CVn variable binary, UX Ari, trends as GJ 436 does, becoming bluer as it becomes dimmer \citep{padmakar:1999,ulvas:2003}. For GJ 436, this behavior does not seem repetative, since during the next dimming phase, the star becomes slightly more red. During the current brightening phase, GJ 436 remains red compared to the 14-year mean color.

UX Ari's behavior of becoming bluer as it gets dimmer is seen both within the binary's orbital period of 6.4 days \citep{carlos:1971} and throughout the 25 year activity cycle \citep{ulvas:2003}. Two explanations for this behavior include flare and facular activity \citep{rodono:1992} or that the relative component of the hotter (and bluer) member of the binary contributes more flux as the cooler one becomes more spotted \citep{mohin:1989,raveendran:1995}. \cite{ulvas:2003b} showed that this behavior can be successfully reproduced using a model that includes dark spots surrounded by bright faculae on the active K star. We thus suggest that GJ 436's odd color behavior as activity changes may be due to the interplay between dark star spots and the faculae that surround them. It may be the case that different activity cycles have different proportions of star spots and faculae. GJ 436's stellar activity cycles be further studied by continuing long-term stellar monitoring of GJ 436 and should be kept in mind when interpreting future high precision transit spectra of GJ 436b.

Since inter-epoch stellar variability and inhomogenous stellar surfaces can induce spurious features in transmission spectroscopy \citep{knutson:2011,fraine:2014,mccullough:2014,oshagh:2013}, it is imperative to check that the brightness variations of GJ~436 do not bias our atmospheric measurements.  In Figure \ref{fig:phot}, we indicate the time and rotational phase of each STIS/G750L and WFC3/G141 transit observation. The WFC3 visits all occur within about a month and so span a range of rotational phases but essentially a single epoch of stellar activity. In contrast, the STIS visits are separated by a year, with the first occurring fairly near stellar minimum, but at otherwise similar magnitude in the rotation period. We discuss the effects of this stellar variability on the transmission spectrum in Section \ref{stellar_activity_effects}.
 
\section{Light Curve Analysis}
 
 \subsection{Limb Darkening}
 
 Crucial to the fitting of transit light curves is proper knowledge of the stellar limb darkening since the effects of limb darkening biases on the calculated transit depth can be on the order of the atmospheric features we expect to measure. Ideally, one could fit for the coefficients that describe the limb darkening \citep{kreidberg:2014}; however, the low signal-to-noise, low temporal sampling, and limited phase coverage of our observations prevents this from being a viable option in our spectral analysis. We thus use stellar models to calculate stellar limb profiles in order to estimate limb darkening coefficients (LDCs). 
 
 We use the ``Limb Darkening Toolkit" (LDTk), which uses the PHOENIX stellar models from \cite{husser:2013} to calculate limb darkening coefficients for a variety of different parameterizations \citep{parviainen:2015}. We use interferometrically determined stellar parameters from \cite{vonbraun:2012}: T$_{eff}=3416$ K, log(g)=4.843, Z=0.02 dex. We also rescaled the $\mu$ values such that $\mu = 0$ occurs where $|\frac{dI}{d\mu}|$ is at maximum as suggested in \cite{espinoza:2015}. We chose the non-linear limb darkening law to fit four coefficients to the stellar limb profile. Since we are fixing the LDCs, it is to our advantage to use more coefficients than in the linear or quadratic limb darkening law, since the resulting fit to the modeled limb intensity profile will be more exact.
 
 We tested LDTk by running a custom PHOENIX model for GJ 436. This has the advantage of using the exact known stellar parameters (Teff, logg, and metallicity) versus interpolating from the grid of models as in LDTk. We found that given the relatively large uncertainties in our data, it made little difference which method was used. We chose to present the LDTk limb darkening coefficients for ease-of-repeatability.
 
 We checked the robustness of the model LDCs by testing LDCs from stellar models  $\pm 100$ K, the current uncertainties for GJ 436's effective temperature. The resulting transit spectra for GJ 436b for the different LDCs are within the errorbars of the spectra using LDCs of the best known stellar parameters. We consider our transit spectra to be robust to uncertainties in the stellar parameters, but caution is warranted as fixed LDCs can lead to biases, especially in the absolute transit depth \citep{espinoza:2015}.
 
  \begin{figure*}[t!]
 	\centering
 	\includegraphics[width=0.98\textwidth]{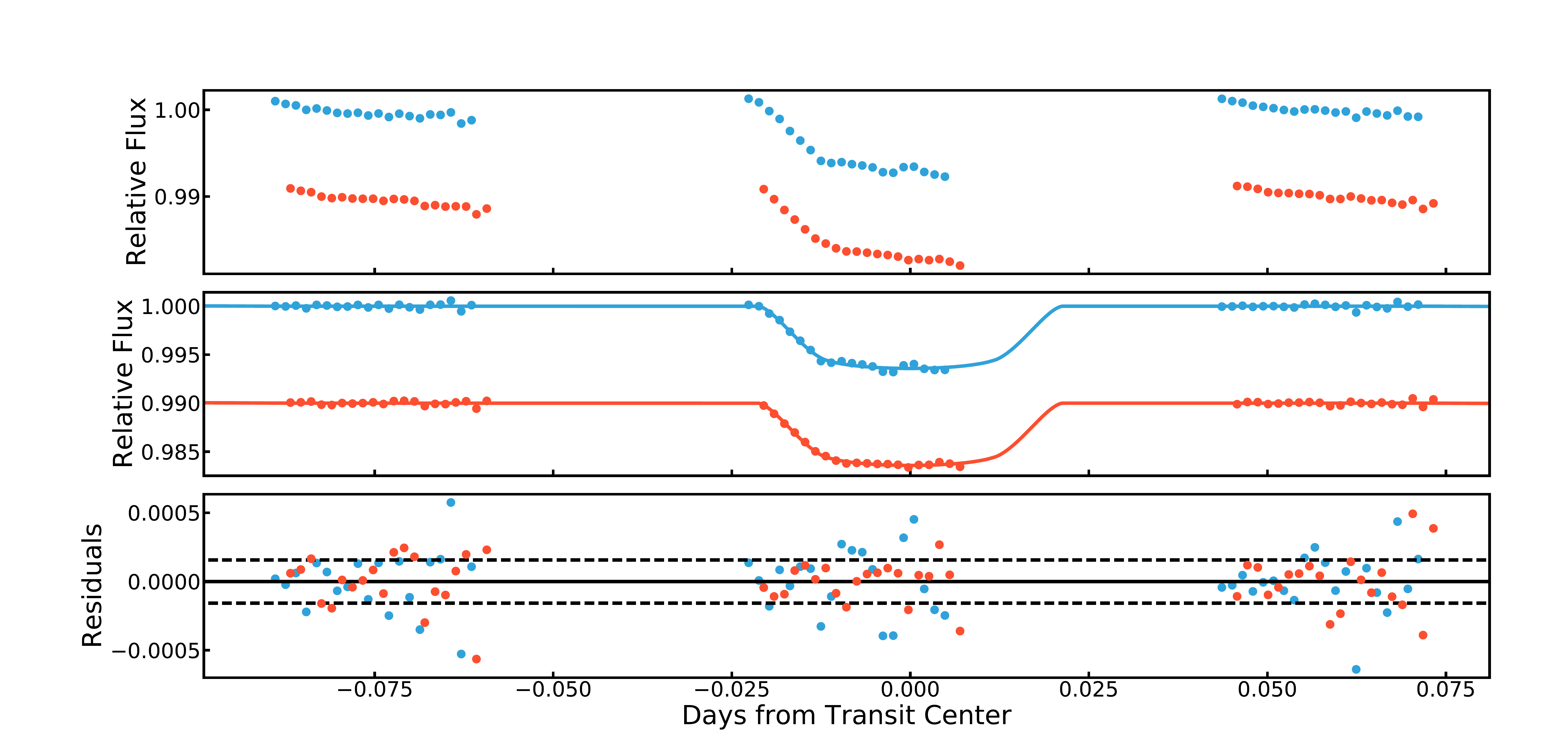}
 	\caption{Top: Raw white light curve normalized to the mean out-of-transit depth. Blue points are those from visit 1, while red are from visit 2 and have been offset by -0.01. Middle: Systematics-corrected white light curve. The solid lines are the respective best-fit models. Bottom: Residuals between the best-fit model and the data. The dashed line represents the expected one sigma uncertainty from photon noise alone. \label{fig:corr_wlc}}
 \end{figure*}
 
 \subsection{Light Curve Model Fitting} \label{fitting}
 
We use the analytical transit model from \cite{mandel:2002} and the transit center, orbital period, inclination, the ratio of the orbital distance to the stellar radius ($a/R_*$), the ratio of the radius of the planet to the radius of the star ($R_p/R_*$) and four parameter non-linear limb darkening coefficients to describe the transit. Due to the observed phasing and the absence of data at the transit's egress, the constraints that our STIS data alone can put on the orbital parameters, particularly the inclination and ratio of orbital distance to the stellar radius ($a/R_*$), are limited. 
We therefore chose to fix these values to the values found in \cite{morello:2015} using PHOENIX limb darkening coefficients. The Independent Component Analysis used in that work has been shown to give repeatable and accurate measurements \citep{ingalls:2016}.
The parameters are listed in Table \ref{params}. The raw and corrected white light curves are shown in Figure \ref{fig:corr_wlc}, along with the resulting residuals.

 \begin{deluxetable}{llccc}
	\tablecolumns{2}
	\tablewidth{0.49\textwidth}
	\tablecaption{Transit Properties Used For GJ 436b  \label{params}}
	\tablehead{\colhead{Parameter} & \colhead{Value}& }
	\startdata   
	Transit Center (T$_c$) (BJD$_{TDB}$) \tablenotemark{a} &  $ 2454222.616632 \pm 0.00012$ \\
	Period (days) \tablenotemark{a}  &  $2.6438986 \pm 0.0000016$ \\
	Inclination (degrees)   &  $86.49 \pm 0.12$\\
	$a/R_*$   &  $13.82 \pm 0.34$ \\
	Impact Parameter (b)         &  $0.846 \pm 0.05$ \\
	Eccentricity \tablenotemark{b}		  & $0.16 \pm 0.02$ \\
	Longitude of Periastron (deg) \tablenotemark{b} &	$351 \pm 1.2$ \\	
	\enddata
	\tablenotetext{a} \protect{From \cite{caceres:2009}.}
	\tablenotetext{b} \protect{From \cite{maness:2007}.}
\end{deluxetable}

\begin{figure}[t!]
 	\begin{center}
 		%\vspace{-20pt}
 		\includegraphics[width=0.5\textwidth]{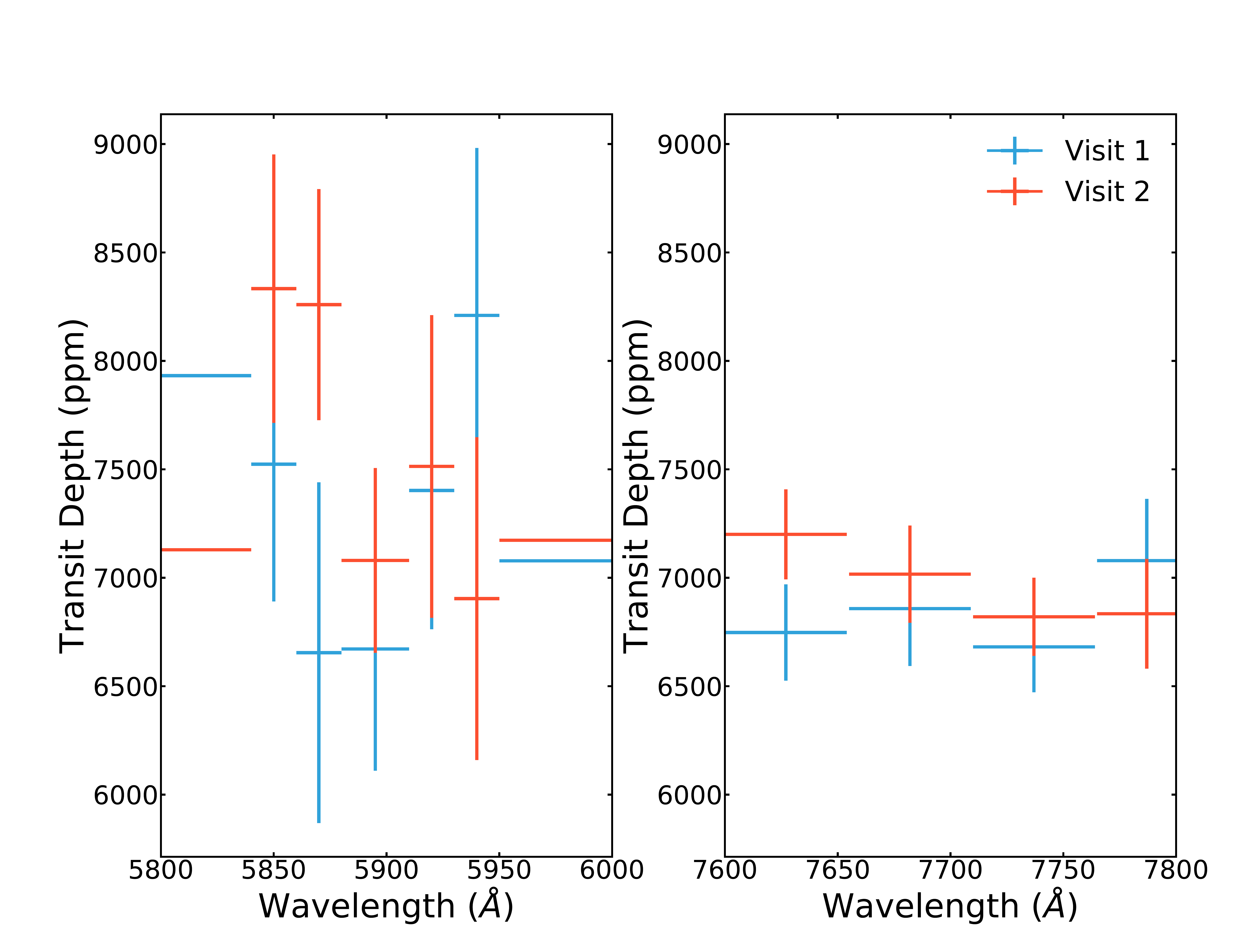}
 	\end{center}
 	%\vspace{-20pt}
 	\caption{Left: Higher resolution binning around the Na absorption doublet ($5890~ \& ~5900$ \AA). Right: Higher resolution binning around the K absorption doublet ($7667 ~\& ~7701$ \AA). Neither visits show any additional absorption at these wavelengths. \label{fig:atomic}}
 	%\vspace{-5pt}
 \end{figure}

The spectra were then split into wavelength bins to see how GJ 436b's transit depth varied as a function wavelength. As mentioned above, we fixed the orbital parameters, allowing only the transit depth and systematic model parameters to vary. Many different wavelength binning schemes were tested and we present here an analysis with 10 bins of approximately 500 \AA \space width. Various numbers of smaller bins and locations were analyzed in a search for absorption from Na or K in the atmosphere of GJ 436b at approximately 5895 and 7684 \AA, respectively (see Figure \ref{fig:atomic}). To fit a transit model to the data as well as estimate uncertainties, we use both Levenberg-Marquardt and Markov Chain Monte Carlo algorithms. Markov Chain Monte Carlo, or MCMC, sampling methods are often used in parameter estimation as they provide empirically estimated uncertainties by exploring the posterior space of parameter likelihoods.  We use the Affine Invariant MCMC sampler implementation {\tt emcee} \citep{foreman-mackey:2012}. We find that both Levenberg-Marquardt and MCMC give very similar results for the parameters and uncertainties estimation. This is likely a result of the gaussianity of the posterior space when orbital parameters and limb darkening coefficients are not free parameters. Because of the number of fits required, we chose to use the Levenberg-Marquardt for the marginalization analysis described in Section \ref{marg}. The average standard deviation of residuals for our bins is about $1.25\times$ the expectation from photon noise alone, similar to other transit spectroscopy studies with STIS \citep{sing:2015}.

 \begin{figure}[t!]
	\center
	\includegraphics[width=0.5\textwidth]{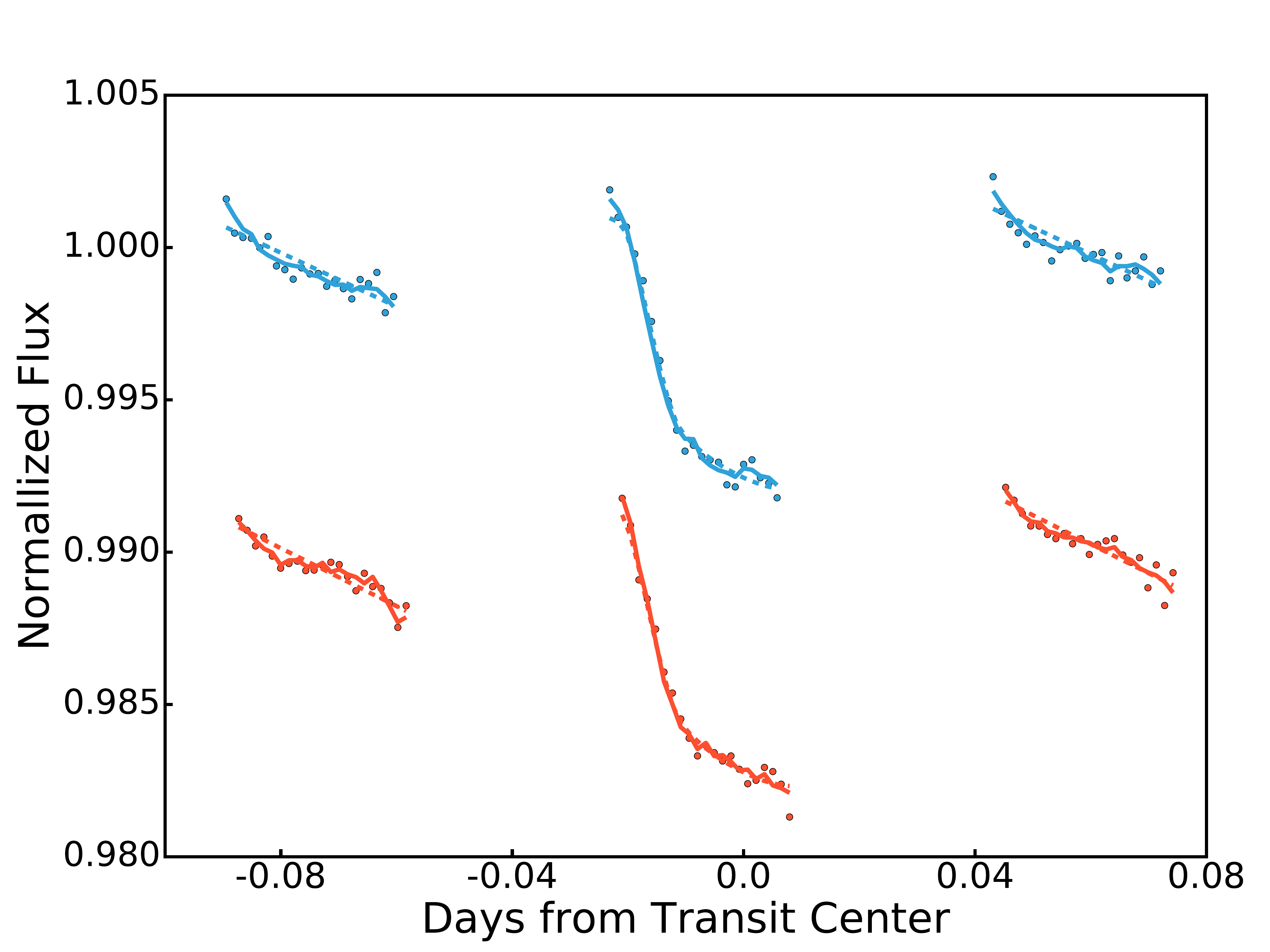}
	\caption{Fits to the raw data for two different models for the systematics. The solid blue line is a fit to the raw data in a wavelength bin from 7721 to 8210 \AA \xspace from visit 1 (blue points) using the most complex systematics model: 4th order polynomial fit to S1, 1st order fit to S2, and 2nd order fit to S3, S4, and S5 (see text for details). The blue dashed line is a fit to the visit 1 raw data only using a 1st order fit to S1 and S2. The red lines and points are the same as above, but for visit 2, offset by -0.01. \label{fig:marg_sys_instances}}
\end{figure}
 
 \begin{figure}[t!]
	\center
	%\vspace{-20pt}
	\includegraphics[width=0.5\textwidth]{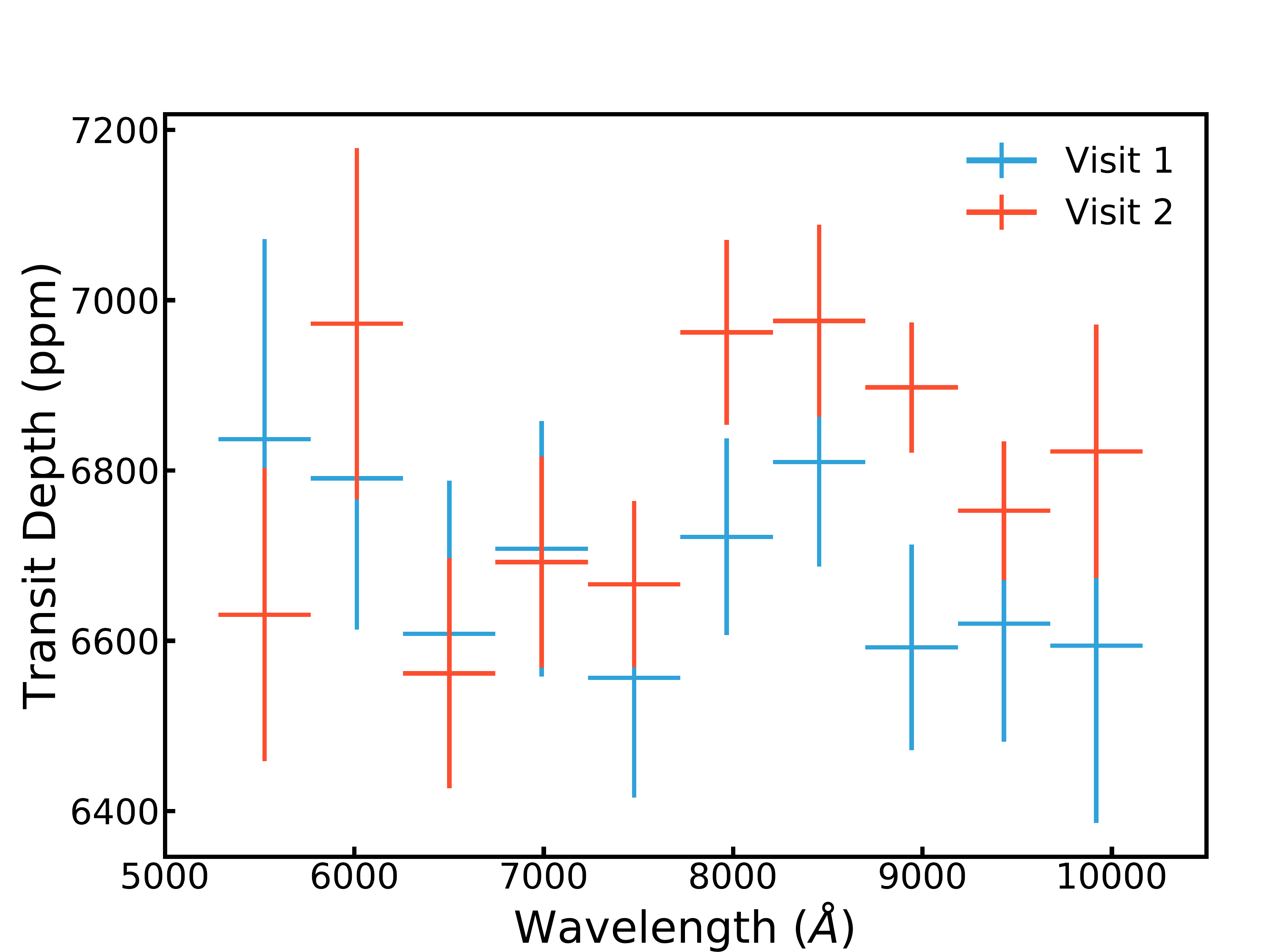}
	%\vspace{-20pt}
	\caption{Transmission spectra of GJ 436b from the individual HST/STIS visits. Each data point is an average of 108 different systematic models weighted by their evidence (see Section \ref{marg}). Both visits are in general agreement. \label{fig:stis_spec}}
	%\vspace{-5pt}
\end{figure}
 
 \subsection{Marginalization} \label{marg}
 
 %Should I make a table of all the systematics??

The top panel of Figure \ref{fig:corr_wlc} shows the effect of instrumental systematics throughout the orbit.  \cite{wakeford:2016} and \cite{sing:2016} describe a method for accounting for systematic uncertainties in HST data by marginalizing over the results from many different systematic models. Instead of choosing a single systematic model as the `correct' model, the full spectral time series analysis is repeated for a variety of systematic models. The measured transit depths are then combined in a way that weights systematic models that describe the data well, while also penalizing models with many free parameters. For this analysis, we choose 5 different covariates to parameterize the systematics found in STIS data: S1) HST orbital phase S2) Time S3) the position of the spectral trace in the spatial direction S4) Slope of the spectral trace S5) the position of the spectral trace in the dispersion direction. S1, S3, S4, and S5 are well-known parameters used to describe systematics caused by HST's low Earth orbit; HST is continually cycling through day-night temperature variations throughout its ~90-minute orbit. This results in small changes in the telescope's focus as components expand and contract \citep{hasan:1994,sing:2013}. %This results in thermal breathing as the telescope expands or contracts and components warm or cool. 
S2 accounts for a visit-long linear change in the measured flux from the star and is similarly thought to be an effect from the telescope itself. S3 and S4 are found through a linear fit to the spectral trace before image rectification. S5 is found through the cross-correlation of the spectrum with a reference spectrum (in this case the fringe flat) in the cross-dispersion direction during 1D spectral extraction.
 
Each covariate is fit with different orders of polynomial: S1 is fit from 1st up to 4th order, which previous STIS studies have shown to adequately capture this trend \citep[e.g.,][]{sing:2013,nikolov:2014}. S2 is kept as a 1st order linear trend, as we find no evidence of a quadratic variation in flux with time and such a variation would be degenerate with the transit model. S3, S4, and S5 are all fit up to 2nd order. As we show below, our analysis and previous analyses have not found it justified to include any higher orders. Each systematic model is then used to fit the data as described in Section \ref{fitting}, resulting in 108 different fits to the data. Figure \ref{fig:marg_sys_instances} shows the difference in the fit using the least versus the most complex systematics model for each visit. The evidence-of-fit, or the marginal likelihood, of each model is calculated in order to compare each of the models. Models that fit the data well will have a high evidence-of-fit and will therefore be given a greater weight in the final marginalization. However, each model will be penalized according to their complexity. This is commonly done with the Bayesian Information Criterion (BIC), or as in Wakeford et al. (2016) with the Akaike Information Criterion (AIC). 
	
	The AIC evidence function we use is

\begin{equation}
\ln E_q = -N \ln \sigma - 0.5 N \ln 2 \pi - 0.5 \chi ^2 - M,
\end{equation}
 
\noindent
 where $E_q$ is the `evidence' for a given systematic model, $N$ is the number of data points being fit, $\sigma$ is the uncertainty placed on the data, $\chi$ is the chi-square statistic and $M$ is the number of free parameters being fit (see Equation 13 of \cite{wakeford:2016}).
 
We find that different systematic models are preferred for each visit. Both visits require systematics models with a third order polynomial of the HST orbit phase (S1). However, the preferred systematics model for visit 1 also includes a second order polynomial to describe the slope of the spectral trace on the detector. This suggests that different systematics may be affecting the data for different STIS visits, even if the target and phasing are the same. The difference in evidence between systematics models with and without the additional spectral trace covariate was not large ($\Delta_{AIC} = 3.5$), however.
 
\begin{figure*}[t!]
	\begin{center}
		\includegraphics[width=.9\textwidth]{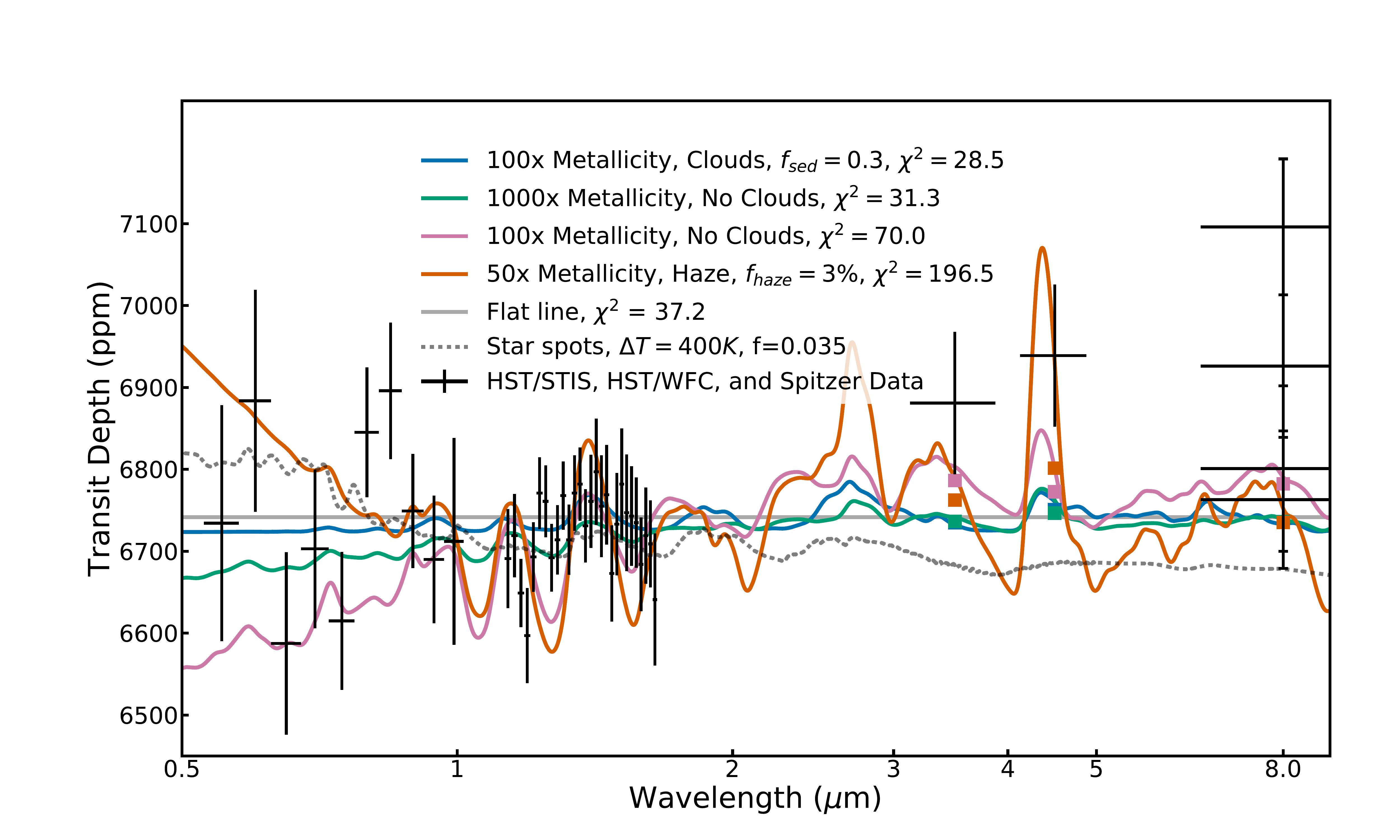}
	\end{center}
	\caption{Combined optical-to-IR transmission spectrum of GJ436b. Each STIS data point (those shortward of 1 \microns) is an average of 108 different systematic models weighted by their evidence (see Section \ref{marg}). Data points from 1.2-1.7 \microns \xspace are from HST/WFC3 \citep{knutson:2014a} and points at 3.6 \microns \xspace and 4.5 \microns \xspace are from \spitzer \xspace \citep[Table 6 in][]{morello:2015}. The 8 \microns \xspace\textit{Spitzer} photometry points are from \cite{knutson:2011}. This transmission spectrum reveals no identifiable features and is consistent with both high metallicity scenarios and moderate metallicity scenarios that include clouds. There are 39 degrees of freedom with respect to the model chi squares. The WFC3 points from 1.1-1.7 \microns \xspace have been offset to match the STIS spectrum (see Section \ref{model_compare}). Note that the wavelength range from approximately 5750-6250 \AA \xspace (second from the left) is affected by wavelength dependent systematics (see Figure \ref{fig:sys_compare}). Also included is the expected slope induced by star spots for a photosphere-star spot temperature contrast of 400 K and a spot coverage of 3.5\%. \label{fig:whole_spec}}
\end{figure*}

 The final result for the marginalized transit depths do not depend on a single systematic model. Since the exact origins of many of the systematic trends seen in HST data are not fully understood, it would be impossible to claim a single `correct' systematic model. The advantage that the marginalized results do not depend on a single systematics model is shared by the more complex Gaussian Process (GPs) technique, but unlike GPs, our method is still a parameterization. Figure \ref{fig:stis_spec} shows the mean transit depth of all 108 systematic models weighted by their evidence for the individual visits. %We discuss the differences between the two visits in Section \ref{results}. 
Figure \ref{fig:whole_spec} and Table \ref{table:specdata} shows the spectra from both visits combined via a weighted average along with several model scenarios and previous observations. We find that the two visits are statistically consistent with each other by calculating the chi-square between the two visits. We note that the long wavelength half of the visit 2 transit spectrum appears consistently above the long wavelength half of the visit 1 spectrum. We show below that this is difference is likely not from stellar variability (see Section \ref{stellar_activity_effects}).

 \begin{deluxetable}{cclll} 
 \tablecolumns{2}
 %\tablewidth{0.45\textwidth}
 \tablecaption{Optical Transmission Spectrum for GJ 436b \label{table:specdata}}
 \tablehead{\colhead{Bin (\AA)} \hphantom{bigspacebig}& \colhead{Transit Depth (ppm)}}
 \startdata   
 			5282-5769 \hphantom{bigspacebig}&  $6734 \pm 139$\\
 			5769-6257   \hphantom{bigspacebig}&  $6884 \pm 128$ \\
 			 6257-6745   \hphantom{bigspacebig}&  $6588 \pm 110$\\
             6745-7233   \hphantom{bigspacebig}&  $6703 \pm 97$\\
             7233-7722   \hphantom{bigspacebig}&  $6615 \pm 84$\\
             7722-8210   \hphantom{bigspacebig}&  $6845 \pm 79$\\
             8210-8698   \hphantom{bigspacebig}&  $6896 \pm 83$\\
             8698-9186   \hphantom{bigspacebig}&  $6749 \pm 69$\\
             9186-9674   \hphantom{bigspacebig}&  $6690 \pm 77$\\
             9674-10162   \hphantom{bigspacebig}&  $6712 \pm 126$\\
 \enddata 
 \end{deluxetable}

 \subsection{Common Mode Systematics}
 
 Another complementary approach to removing systematics involves correcting for the common-mode wavelength independent-systematics before the removal of wavelength dependent systematics \citep{sing:2016,nikolov:2014,huitson:2012}. The common-mode systematics are found by fitting a full systematics model to the white light curve data and then are `divided out' of the individual spectral bins. The spectral bins are theoretically left with only the wavelength-dependent systematics. This method can allow for the fits to the individual spectral bins to require fewer free parameters, helping to decrease the uncertainty in $R_p/R_*$. We implemented the common mode correction to this analysis, but found it did not significantly improve either the fits or the estimated uncertainties. The common mode correction did improve the evidence for less complex systematics models, but none increased the evidence enough to justify their use over more complex systematics models.
 
\begin{figure}[t!]
	\center    \includegraphics[width=3.5in]{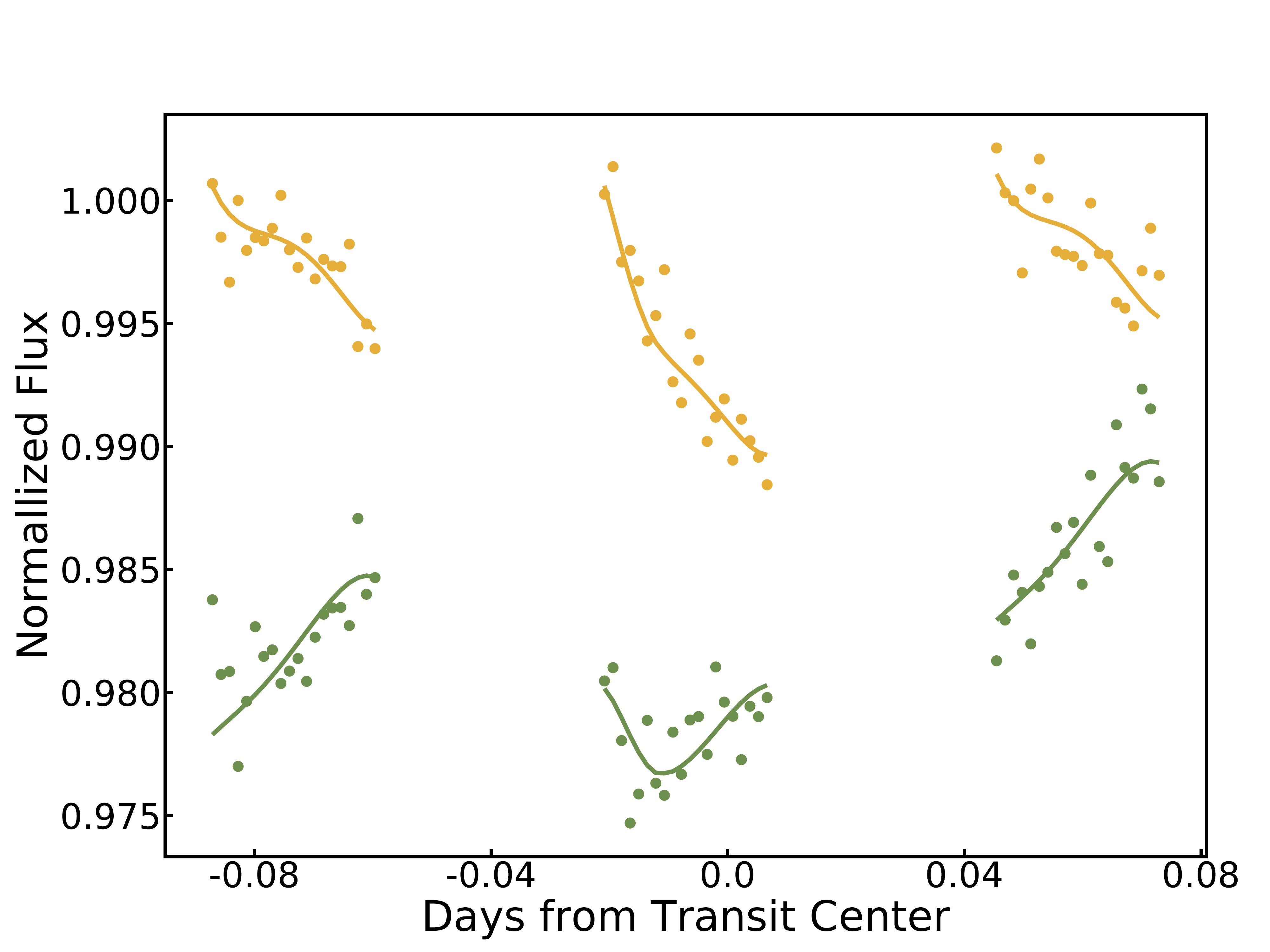}
	\caption{The yellow points are the measured fluxes from a small bin from 5820-5840 \AA \xspace for visit 2. The systematics exhibited in this wavelength bin are representative of those seen across the majority of the detector. In contrast, the green points are from an adjacent bin, 5840-5860 \AA, and exhibit a very different systematic trend. The solid lines are the respective best-fit transit models. \label{fig:sys_compare}}
\end{figure}

It was found that some areas of the detector exhibited different systematic trends than the rest of the detector. Figure \ref{fig:sys_compare} shows the difference in the measured raw flux from two adjacent 20 \AA \space bins near the stellar Na line during visit 2. While still being described by the same systematics model, the overall trends are different, especially in those systematics associated with HST's orbital phase. Instead of decreasing in flux throughout the orbit, some bins increase throughout the orbit. The same effect is seen in visit 1. Areas like this on the detector will prevent the application of the common mode systematic technique from being beneficial. 

An explanation for this effect could be uncorrected systematics in the position of the spectral trace. Since the bins that exhibit different systematics tend to be those in which the stellar spectrum is decreasing with increasing wavelength at the bin edges, shifts in the spectral trace can lead to strong and differing systematics in adjacent bins. However, even after applying the dispersion solution and shifting the spectra to a common rest frame via cross correlation, this effect is still present.
 
 \section{Results} \label{results}

 We varied bin sizes and locations in order to isolate any signal from Na and K, but found no significant absorption in either case (see Figure \ref{fig:atomic}). This finding is well within theoretical expectations as models do not predict significant absorption of either Na or K in transit if the atmosphere is even moderately cloudy or high metallicity (see Figure \ref{fig:whole_spec}). Because of GJ 436b's low equilibrium temperature,  sodium and potassium may be sequestered in clouds of Na$_2$S and KCl, suggesting that the pressures probed by our observations are cooler than the condensation temperature of Na$_2$S and KCl ($<\sim 700$ K). It may be these Na$_2$S and KCl clouds that are muting the 1.4 \micron \xspace water absorption feature. Additionally, a detection of Na with STIS for Gj 436b is especially difficult due to the low flux from the M-dwarf host star at these shorter wavelengths.

While a scattering slope may still be present at wavelengths shortward of $\sim$ 0.6 $\mu m$, we do not detect a significant slope in the STIS G750L passband. In fact, given the spectrum's agreement with a flat line, we do not detect GJ 436b's atmosphere. This is consistent with the interpretation that clouds or high-metallicity could be muting spectral features in transmission. We discuss below the effect that stellar activity can have on the spectrum (see Section \ref{stellar_activity_effects}). We also quantify the offsets resulting from different orbital solutions (see Section \ref{orbital_solution_effects}).
 
 \subsection{The Effect of Stellar Variability} \label{stellar_activity_effects}
 
Since unocculted star spots and plages can cause slopes at optical wavelengths, we investigate the role stellar activity may play in GJ 436b's transit spectrum \citep{berta:2011,mccullough:2014,rackham:2017}. The difference between the flux from a stellar photosphere and the flux from a star spot is a wavelength dependent quantity since star spots are redder due to their lower temperature. When a planet transits the stellar disk, a slope will be induced in the spectrum because unocculted star spots will contribute a greater proportion of the observed stellar flux at shorter wavelengths, making the planet appear larger at short wavelengths. Stellar plages will have a similar but opposite effect on the transit spectrum.
 
\cite{mccullough:2014} analyzed the effect of unocculted starspots on the spectrum of HD 189733b, concluding that they could explain the observed spectral slope as well as Rayleigh scattering. The apparent transit depth as a function of wavelength may be modeled as:

\begin{equation}
\frac{\tilde{R_p^2}}{\tilde{R_s^2}} = \frac{R_p^2}{R_s^2} \frac{1}{1-\delta(1-F_{\nu}(spot)/F_{\nu}(phot))}
\end{equation}

where $\tilde{R_p^2}/{\tilde{R_s^2}}$ is the observed transit depth, ${R_p^2}/{R_s^2}$ is the actual radius ratio squared, $\delta$ is the fraction of the star's projected surface area that contains spots, and $F_{\nu}(spot)/F_{\nu}(phot)$ is the ratio of flux from the star spot to the flux from the photosphere \citep{mccullough:2014}. Since the optical spectrum of M dwarfs can be rich in opacity features, we model $F_{\nu}(phot)$ and $F_{\nu}(spot)$ using PHOENIX atmosphere models rather than approximating them as blackbodies. Given the results of our photometric stellar monitoring, we estimate below the effect that stellar activity could have on GJ 436b's transmission spectrum.

The maximum difference in stellar flux from our monitoring of GJ 436 is 13 mmag ($\sim $ 10 mmag from the stellar activity cycle and 3 mmag from rotational variability). A difference in stellar magnitude of 13 mmag corresponds to a difference in flux of 1.4\%. This flux difference corresponds to variability in the star spot coverage up to  6.3\% for photosphere-starspot temperature constrasts of 200 K (T$_{phot} = 3400$ and T$_{spot} = 3200$) and 3.5\% for contrasts of 400 K (T$_{spot} = 3000$). With these values, we can expect slopes of up to 150 ppm to vary throughout the activity cycle, as seen in Figure \ref{fig:whole_spec}, if star spots are the main source of stellar variability. If plages are dominating the stellar variability, the slopes induced would be opposite those from star spots (i.e., transit depths would be smaller at shorter wavelengths). Note that, as mentioned above, GJ 436 does not exhibit the expected color variation if star spots dominate the variability and the interplay between star spots and steller plages will complicate this picture.

We mention above that the two STIS visits gave statistically consistent spectra, yet appeared to show some correlation in that the long wavelength half of the visit 2 transit spectrum showed larger transit depths than the long wavelength half of the visit 1 spectrum (see Section \ref{marg}). If this were due to stellar plages inducing a slope toward smaller transit depths at shorter wavelengths in the spectrum of visit 2, we would expect the visit 1 transit depths to be significantly higher at all wavelengths. Stellar plages only serve to reduce transit depths relative to a homogenous stellar surface with the same photospheric temperature. Alternatively, if star spots were affecting visit 1, reducing the long wavelength part of the visit 1 transit spectrum relative to visit 2, we would expect the blue part of the spectrum to diverge even more. In general, if an inhomogenous stellar photosphere were causing any difference between the two STIS visits, we would expect the long wavelength half of the spectra to agree better relative to the short wavelength half. We therefore conclude that any difference between visit 1 and visit 2 appears not to be due to stellar variability.
 
\subsection{Effects of Different Orbital Solutions} \label{orbital_solution_effects}

\begin{figure}[t!]
 	\begin{center}
 		\includegraphics[width=0.5\textwidth]{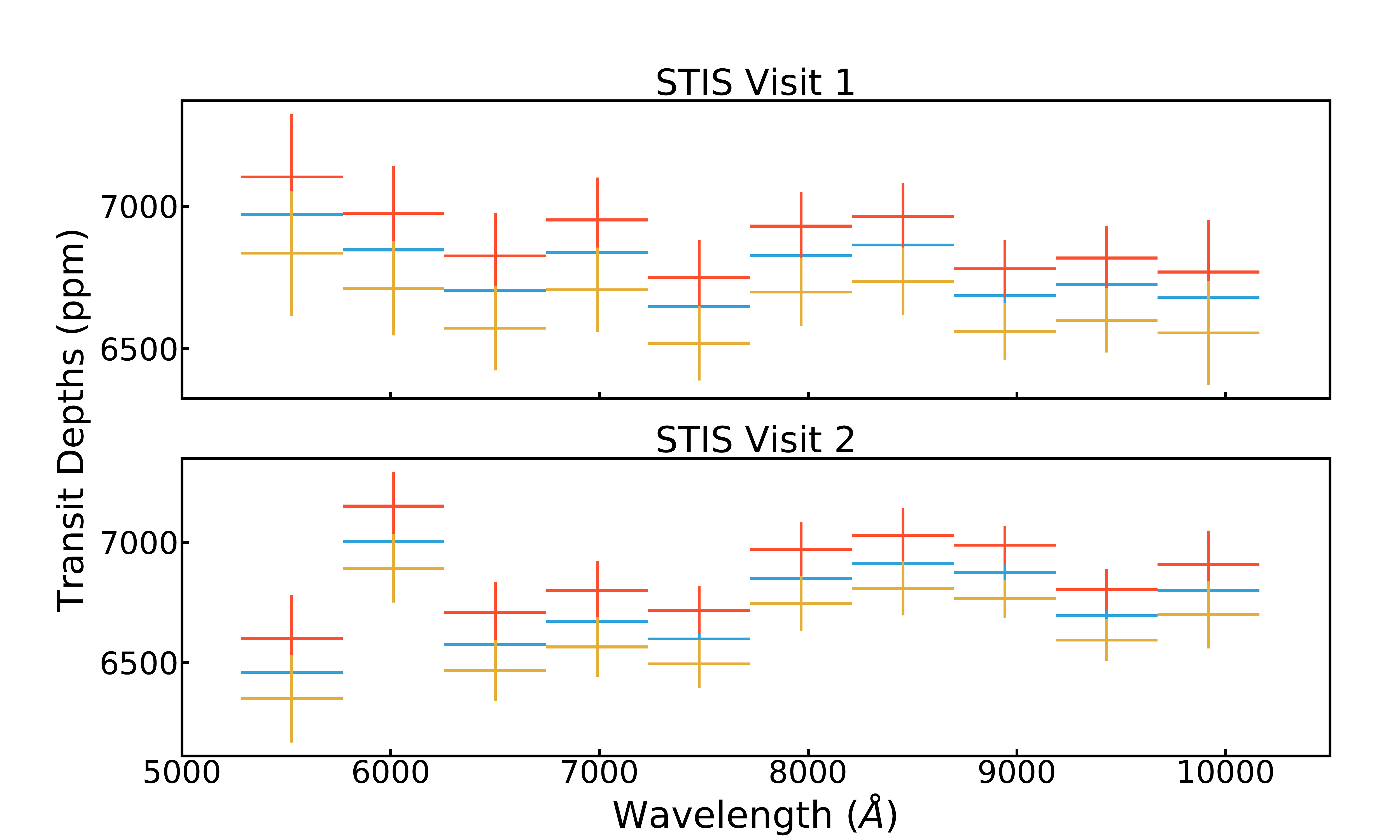}
 	\end{center}
 	\caption{Top: Resulting transmission spectrum for visit 1 when using different orbital solutions. Bottom: Same, but for visit 2. Blue points correspond to values from Table 6 of \cite{morello:2015}, yellow corresponds to \cite{knutson:2014a}, and red corresponds to \cite{lanotte:2014}. See Table \ref{table:orbs} for the orbital parameters used. \label{fig:orbit_compare}}
 \end{figure}

Since we chose to fit the orbital solution to values found in the literature, we compared orbital solutions from \cite{knutson:2014a}, \cite{lanotte:2014}, and \cite{morello:2015} (see Table \ref{table:orbs}). Figure \ref{fig:orbit_compare} shows how the resulting STIS spectrum of GJ 436b compare. For each orbital solution, there is a uniform offset. The largest difference is between the \cite{knutson:2014a} and \cite{lanotte:2014} orbital solutions, with an average offset of about 260 ppm. Because the offset is uniform in wavelength, we can be confident that the orbital solution does not affect our non-detection of both a scattering slope and alkali features.

 \begin{table*}[t] 
\centering  
 \caption{Orbital Solutions for GJ 436b}
\label{table:orbs} 
\begin{tabular}{p{3.2cm}p{1.8cm}p{1.8cm}p{2.5cm}p{2.2cm}p{2.2cm}}
\hline
Reference & inclination (degrees) & $a/R_*$ & impact parameter (b) & $e$ & $\omega$ \\
\hline 

 			\cite{morello:2015}  &   $86.49 \pm 0.12 $&$ 13.82 \pm 0.34 $& $0.846 \substack{+0.050 \\ -0.049} $& $0.16 \pm 0.02$ & $351 \pm 1.2$ \\
 			 \cite{knutson:2014a}   &    $86.774 \pm 0.03 $& $14.41 \pm 0.10 $& $0.81 \substack{+0.014 \\ -0.012}$ & $0.1495 \pm 0.016$ & $336 \pm 12$ \\
 		\cite{lanotte:2014} &  $86.858 \substack{+0.049 \\ -0.052}$& $14.54 \substack{+0.14 \\ -0.15}$& $0.7969 \pm 0.021$ & $0.1616 \pm 0.004 $ & $327.2 \pm 2.2$ \\
 \end{tabular}
 \end{table*}
  
\subsection{Model Comparisons} \label{model_compare}
 
We compare the optical-to-infrared spectrum of GJ 436b to the models described in \cite{morley:2016}. These models are self-consistent radiative-convective models in both chemical equilibrium and disequilibrium via quenching. In the models with quenching, the  abundances of CH$_4$, CO, and CO$_2$ at pressures below 10 bars are fixed to be the equilibrium abundances at 10 bars (i.e., the quench pressure is 10 bars). Interior heating is included from tidal dissipation because of GJ 436b's eccentric orbit \citep[$e\sim 0.15$,][]{turner:2016}. Clouds are modeled through a modified version of the \cite{ackerman:2001} models \citep{morley:2012,morley:2013,morley:2015}. In this cloud model, $f_{sed}$ represents a sedimentation parameter that determines whether the clouds are thin ($f_{sed} > \sim0.3$) or thick  ($f_{sed} < \sim0.3$). Photochemical hazes are modeled using the results from \cite{line:2011} with the approach from \cite{morley:2013,morley:2015}. This method models the creation of soot precursors (C$_2$H$_2$, C$_2$H$_4$, C$_2$H$_6$, C$_4$H$_2$, and HCN) and `converts' them to scattering hazes via a haze efficiency parameter, \fhaze \xspace (i.e., \fhaze \xspace is the fraction of soot precursors that actually form hazes)  \citep{line:2011}.

We allowed for a uniform offset between the STIS spectrum and the WFC3 spectrum in our calculation of the $\chi ^2$ values. The original white light curve depths found with the \cite{knutson:2014a} orbital solution was 7000 ppm. This is above the average transit depth of the STIS spectrum of around 6700 ppm using the \cite{morello:2015} orbit. Since the overall offset between the modeled optical and infrared spectrum is small, we choose to implement a simple offset.. Because the WFC3 spectrum was analyzed using a template fitting technique, the differential depths within the spectrum would not change given a uniform offset.

We find that the best-fit model from \cite{morley:2016} (1000x metallicity, T$_{int}$ = 240 K, $f_{sed}$ = 0.3 salt/sulfide clouds and disequilibrium chemistry through quenching of CH$_4$, CO, and CO$_2$) is consistent with our new HST/STIS observations. This model has a moderately thin cloud layer of salt/sulfide grains. Lower $f_{sed}$ ($\thicksim$0.1) can induce a slope in the transmission spectrum toward lower transit depths at shorter wavelengths, but our observations do not have the sensitivity to constrain this well. As in \cite{morley:2016}, clear high metallicity models ($\sim 1000$ x solar) can explain the full transmission spectrum as well as models with lower metallicity but with some clouds (e.g., $\sim 100$ x solar metallicity + 0.3 $f_{sed}$). Since both scenarios are relatively flat in the optical, they are both consistent with our observations. We must look to future observations in the infrared to determine which scenario is taking place on GJ 436b (see Sec \ref{jwst}).

Our observations help rule out most hazes with low \fhaze \xspace efficiency. The observed transit depths around 1.4 \microns \xspace are inconsistent with models having large mean haze particle radii and low \fhaze values. Our new STIS observations disfavor smaller mean particle radius hazes as well, indicated by the lack of a strong scattering slope in the STIS bandpass (see Figure \ref{fig:whole_spec}). Some haze models are still consistent with the data, but require a large \fhaze \xspace value to sufficiently flatten the transit spectrum.

 \section{Discussion}
 
 \subsection{Comparison with Other Sub-Jovian Exoplanets}
 
Only about a dozen sub-Jovian exoplanets have been characterized spectroscopically in transmission. While the smallest exoplanets have either flat spectra or very tentative detections in their transmission spectra \citep{knutson:2014b,tsiaras:2016}, a few exoplanets closer to the size of the Solar System's ice giants have detectable water absorption features in their NIR spectrum with WFC3. HAT-P-11b and HAT-P-26b have a relatively strong detection of water at 1.4 \microns \xspace \citep{fraine:2014,wakeford:2017b}, while observations of GJ 436b have not yet revealed a water absorption feature \citep{knutson:2014a}. \cite{stevenson:2015} also interpreted 0.7 to 1.0 \microns \xspace ground based observations of HAT-P-26b to show evidence of water. Differences in the amplitude and existence of the 1.4 \micron \xspace water absorption feature could point to variations in either the water abundance or in aerosol properties. \cite{crossfield:2017} provide evidence for a correlation between the amplitude of the water absorption feature and either the equilibrium temperature or hydrogen mass fraction. To explain this correlation, they suggest that low temperature planets may have more optically-thick photochemically produced hazes or that smaller planets have higher metallicity and thus reduced scale heights.

Recently, \cite{wakeford:2017b} constrained the metallicity of the Neptune-sized HAT-P-26b to be $4.8\substack{+21.5 \\ -4.0} \times$ solar using the retrieved water abundance from the 1.4 \micron \xspace feature. This is very different from the retrieved metallicty for GJ 436b from \cite{morley:2016} who found the metallicity to be greater than 106 $\times$ solar using both the featureless IR transmission spectrum and the dayside spectrum. Large differences in metallicities between these planets may point to differences in their formation, perhaps in formation location or disk properties. In agreement with modeling predictions and population studies, the characterization of individual sub-Jovian exoplanets already indicates that this population is diverse in composition \citep{moses:2013,fortney:2013,wolfgang:2016,venturini:2016}.

GJ 436b and HAT-P-26b both have STIS spectra that are interpreted as being cloudy and featureless; however, the STIS spectra are strikingly similar, especially regarding an abrupt change in transit depth at about 0.8 \microns.
This jump is also seen in ground based data of HAT-P-26b taken at \textit{Magellan} with LDSS3-C \citep{stevenson:2015}. The transit depth of HAT-P-26b near 0.9 \microns \xspace is confirmed by WFC/G102 observations from \cite{wakeford:2017b}. Additionally, ground-based data from \cite{rackham:2017} also found a similar slope at comparable wavelengths, this time in GJ 1214b at \textit{Magellan} with IMACS. GJ 1214b's optical spectrum is well fit by a model for the effects of unocculted stellar plages on the photosphere of the host star covering 3.2\% of the stellar disk with a temperature contrast of $\sim$~350 K. We explore this explanation for the other planets below. Table \ref{table:subjov} lists the amplitude, number of transits, host star type, and references for the sub-Jovian planets that show this jump in transit depth. In total, 11 transits of 4 sub-Jovian planets observed from 4 different instruments from 2 facilities have shown this increase in transit depth at 0.8 \microns.

We quantify the significance of this jump in transit depth by comparing $\chi ^2$ fits to a flat line with and without including the data between 0.7 and 0.9 \microns. The $\chi ^2$ between a flat line and the GJ 436b and HAT-P-26b STIS data is 55.5 using 17 data points, giving a reduced $\chi ^2$ of 3.47, indicating that the probability that our data could have come from a flat line is less than 0.05\%. If we remove all data points between 0.7 and 0.9 $\mu$m, we get a $\chi ^2$ of 8.73 using 11 data points, resulting in a reduced $\chi ^2$ of 0.87, indicating that the probability that this data could have come from a flat line is greater than 50\%. This simple analysis serves to show that, when taken together, the STIS spectrum of these planets are inconsistent with a flat line primarily due to a jump in transit depth between 0.7 and 0.9 \microns. In the event that the errorbars are underestimated, we find that they would need to be increased by 35\% for the STIS data to be consistent with a flat line. We note, however, that when we include the ground-based and WFC3 data, we find that a flat line is inconsistent with the data, even when excluding the 0.7-0.9 \microns \xspace region. This makes sense because the GJ 1214b spectrum is far from flat (stellar plages are likely making short wavelength transit depths lower than long wavelength transit depths) and HAT-P-26b may have water absorption around 0.9 microns, moving the ground based and WFC3 data away from a flat line (though this water absorption does not explain the jump in transit depth between 0.7-0.9 microns).

\begin{figure}[t!]
	\begin{center}
		%\vspace{-20pt}
		\includegraphics[width=0.5\textwidth]{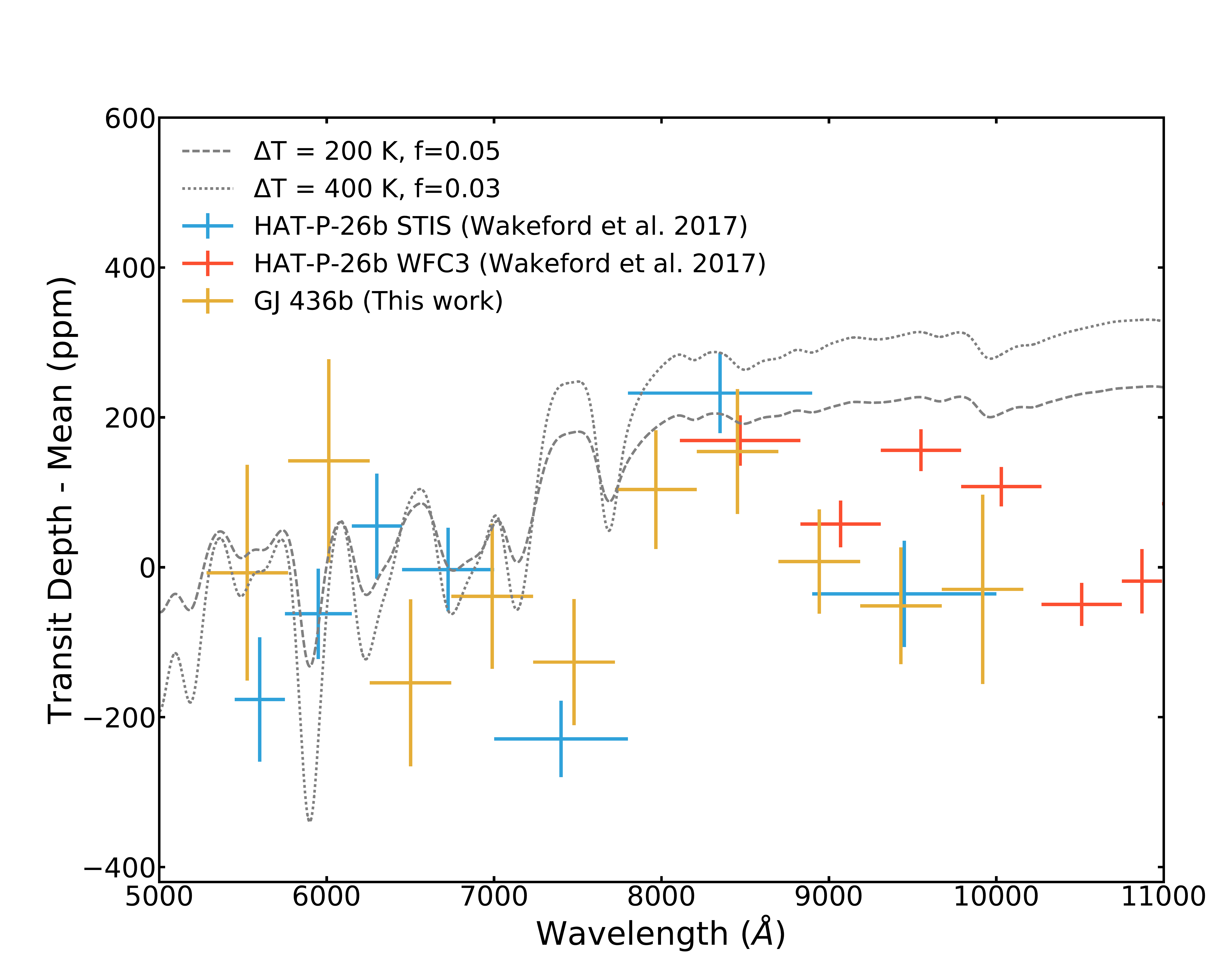}
	\end{center}
	%\vspace{-20pt}
	\caption{HST/STIS G750L transmission spectra of HAT-P-26b and GJ 436b. HAT-P-26b observations from \cite{wakeford:2017b} consist of a single visit with both STIS and WFC3/G102 and our GJ 436b observations consist of two visits with STIS. Both data sets show a significant and sudden rise in transit depth near 0.8 \microns. Similar trends are seen in the ground-based data for HAT-P-26b \citep{stevenson:2015} and GJ 1214b \citep{rackham:2017}, though we omit this data in this figure for clarity. Overplotted are two scenarios for stellar plages. The dashed line corresponds to a plage coverage over the stellar disk of 5\% with a 3000 K photosphere and 3200 K plages. The dotted line corresponds to a plage coverage of 3\% with a 3000 K photosphere and 3400 K plages. The models assume a planetary transit depth of 7000 ppm. \label{fig:subjov}}
	%\vspace{-5pt}
\end{figure}

The explanation of stellar plages is plausible for GJ 436b. GJ 436 is a M-dwarf like GJ 1214b and the maximum flux difference between the photosphere and a stellar plage with a temperature contrast of 350 K is around 0.75 \microns \xspace, where we see the jump. In Figure \ref{fig:subjov}, we plot two models of the effect of stellar plages using equation 11 of \cite{rackham:2017}:

\begin{equation}
\left(\frac{R_p}{R_s}\right)_{\lambda,obs} = \sqrt{1-\frac{(1-f-D_{\lambda})S_o+fS_u}{(1-f)S_o+fS_u}}
\end{equation}

where $f$ is the fraction of the stellar disk with stellar plages, $D_{\lambda}$ is the transit depth expected if the stellar disk had no stellar plages (i.e., the `true' transit depth), $S_u$ is the flux from a stellar plage, and $S_o$ is the flux typical of the greater photosphere.  We use PHOENIX stellar atmosphere models to model the spectra of the photospheres and the hotter stellar plages.

In general, the stellar plage models do not match the jump at 0.8 \microns \xspace well. Longward of 0.8 \microns \xspace, transit depth decreases, unlike expectations from the plage models. Similarly, shortward of 0.7 \microns, the transit depth does not stay low as in the plage models. The explanation of stellar plages also is not a sufficient explanation for HAT-P-26b, whose host star is of type K0. For temperature contrasts of around 300 K, the effects of stellar plages for a K0 photosphere are flat until about 6500 \AA, outside the range of the jump in transit depth. Additionally, the increase in transit depth is rather sudden and does not qualitatively match the trend seen in GJ 1214b that is explained by stellar plages.

Another explanation could be biases from fixing the limb darkening coefficients (LDCs) while fitting the transit light curve. However, \cite{stevenson:2015} tested fitting LDCs and found the results to be consistent with LDCs from models. Additionally, \cite{rackham:2017} left one of the quadratic LDCs as a free parameter, further verifying that GJ 1214b's increase in transit depth is not related to biases in limb darkening. This does remain an explanation worth futher investigation for the other planets.

A third explanation could be an additional opacity source around 0.8 \microns \xspace in the atmosphere of these planets. However, in our grid, no model seems to fit this part of the spectrum accurately and all of the best fitting models are featureless throughout STIS wavelengths. Any additional opacity source at 0.8 \microns \xspace would also need to make sense in context of the rest of the spectrum, which is either featureless or only shows water absorption. Additionally, the size of such a feature would be unprecedented in sub-Jovian exoplanets. The jump in transit depth at 0.8 \microns \xspace in HAT-P-26b is about the size of its water absorption feature at 1.4 \microns. As Table \ref{table:subjov} shows, the jump in transit depth spans up to 5.5 scale heights. Thus we recommend caution before interpreting the 0.8 \micron \xspace feature as an opacity source. We also note that this effect is not readily apparent in the optical spectrum of hot Jupiters \citep[e.g.,][]{sing:2016}.

\begin{table*}[t] 
\centering  
 \caption{Comparison of 7000-8000 \AA \xspace Feature in Warm Neptunes}
\label{table:subjov} 
\begin{tabular}{p{2.2cm}p{2.5cm}p{2.5cm}p{1.5cm}p{1.5cm}p{3.5cm}}
\hline
Planet & Approx. Depth of Feature (ppm) & Approx. Depth of Feature ($\delta R_p / H$) & Number of Transits & Host Star Type & Reference\\
\hline
  
 			GJ 436b &    250 & 5.5 & 2 & M2.5V  & This work \\
 			 GJ 1214b   &    450 & 2 & 3 & M4.5V  & \cite{rackham:2017} \\
 		HAT-P-26b &  500 & 5 & 3 & K0V & \cite{stevenson:2015,wakeford:2017b}
 \end{tabular}
 \end{table*}
  
\subsection{Future Prospects with JWST for GJ 436b's Transmission Spectrum}\label{jwst}

The wide wavelength coverage and superior light gathering capability of the James Webb Space Telescope (JWST) will allow planets like GJ 436b to be studied in more detail. As mentioned above, current JWST GTO plans include GJ 436b eclipse observations in three instruments' GTO programs. These observations will tell us a great deal about the composition and structure of the dayside atmosphere; however, observations of GJ 436b's transmission spectrum are necessary to probe a different region of the planet, namely the terminator. To what degree the dayside atmosphere's composition and structure differ from that of the terminator has important implications for our understanding of the global circulation taking place in these atmospheres \citep{kataria:2016}.
	
\begin{figure}[t!]
	\begin{center}
		\includegraphics[width=0.5\textwidth]{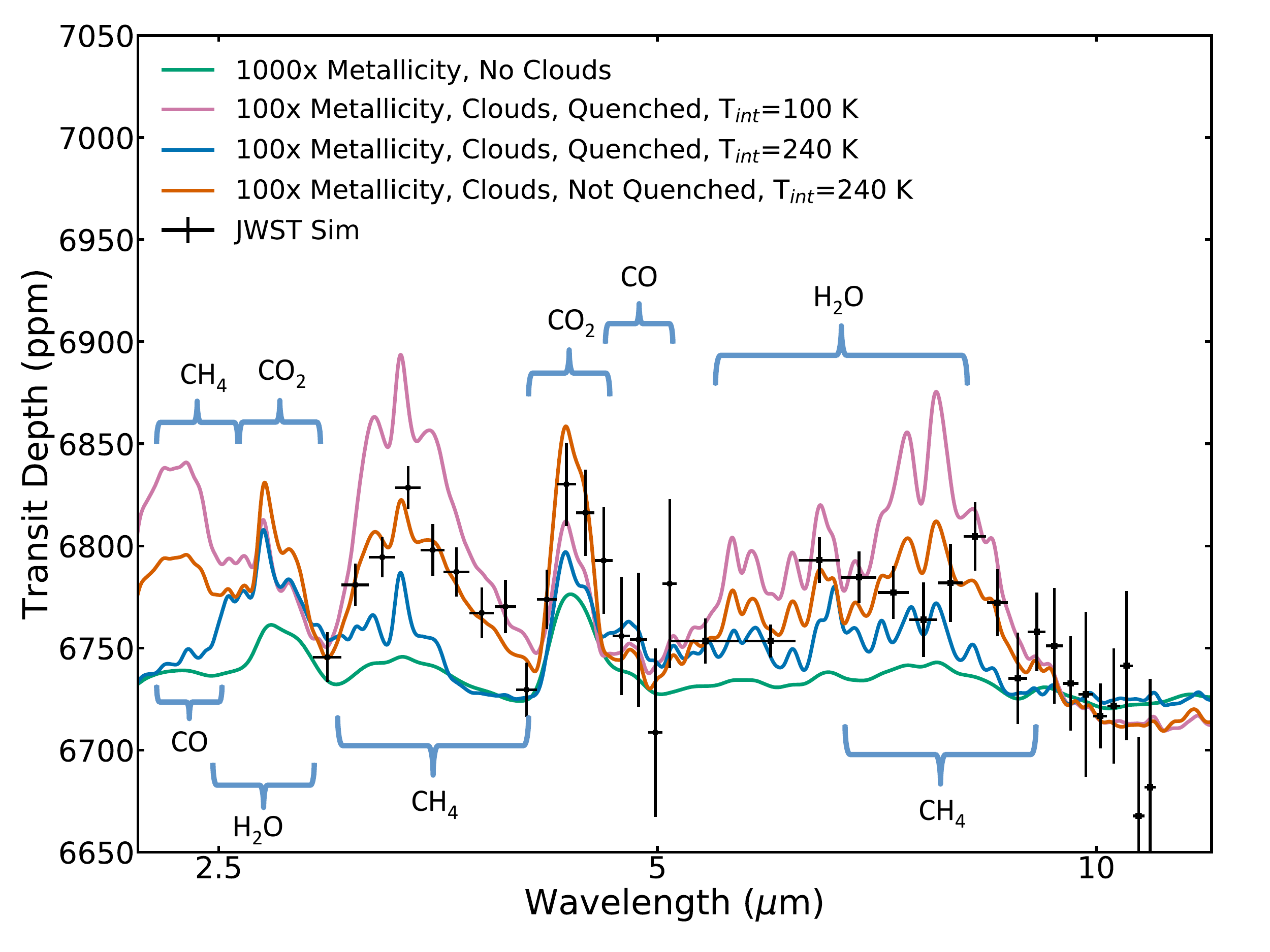}
	\end{center}
	\caption{\small Four model scenarios for the transmission spectrum of GJ 436b. Though each of these scenarios is difficult to discern using HST/STIS and HST/WFC3, observations using JWST will allow us to distinguish between such scenarios. \label{fig:jwst_model}}
\end{figure}

We simulated JWST data using PandExo for 3 transits observed with Near Infrared Spectrograph (NIRSpec) G395H grism Bright Object Time Series and Mid-Infrared Instrument (MIRI) slitless Low Resolution Spectroscopy with a noise floor of 10 ppm \citep{batalha:2017}. We avoided shorter wavelength instrument modes that may have trouble with saturation and non-linearity effects due to the brightness of GJ 436. The simulated data are shown in Figure \ref{fig:jwst_model} with 200 and 10 pixels per bin for NIRSpec and MIRI, respectively. This is comparable to a resolution of R$\sim$20-40 for NIRSpec and R$\sim$10-55 for MIRI. We include four model scenarios from \cite{morley:2016} and described in Section \ref{model_compare} for GJ 436b's transmission spectrum at JWST wavelengths. For the simulated data, we adopted the 100 $\times$ metallicity scenario with internal heating but no quenching of CH$_4$, CO, and CO$_2$ in order to demonstrate JWST's ability to distinguish between model scenarios.

All four model scenarios in Figure \ref{fig:jwst_model} are consistent with current measurements. However, JWST transit observations can distinguish cloudy and high-metallicity scenarios, quenched and non-quenched scenarios, and different internal heating. The data point uncertainties approach the imposed noise floor of 10 ppm, but since the spectral features in the transit spectrum have amplitudes much larger than the noise floor, each of the models could be distinguished even if the noise floor were increased. \cite{morley:2016} found that quenched carbon chemistry with internal heating provided the best fit to the dayside thermal emission spectrum. With transit spectra from JWST, we can test this conclusion at the planet's terminator. Additionally, JWST will be capable of resolving the spectral features of CH$_4$, CO, and CO$_2$ in both transmission and thermal emission spectra, a big step forward from the \spitzer \xspace photometry used today to infer the presence of these molecules.

It is worth noting that JWST's short wavelength capabilities end at about 0.7 $\mu m$, meaning that space-based optical and UV observations with STIS and ground-based optical and near-UV observations (e.g.,\ the Arizona-CfA-C\'atolica Exoplanet Spectroscopy Survey (ACCESS) and the GTC exoplanet transit spectroscopy survey) will remain a useful and unique way to characterize these exoplanets. However, as we have seen, properly interpreting the optical transmission of an exoplanet requires an accurate understanding of effects from the host star.
 
 \section{Conclusion}
 
We presented new HST/STIS transit observations that help to constrain the sources of opacity in the atmosphere of GJ 436b. Even though our observations cannot distinguish between very high metallicity scenarios and lower metallicity scenarios with clouds, we are able to show that both scattering from thin hazes (low $f_{haze}$) and low metallicity cloud-free scenarios are inconsistent with current data. These conclusions agree with those from \cite{morley:2016}.

We also found a strong similarity between GJ 436b's optical spectrum and those from two other sub-Jovian exoplanets, especially regarding an abrupt increase in transit depth near 0.8 \microns. The effects of stellar plages can explain the optical spectrum of GJ 1214b, but does not explain the rather sudden change in transit depth for the other planets, especially HAT-P-26b. While we have not ascribed a cause for this feature, future characterization and modeling will ultimately shed light on this problem.

Our stellar photometric monitoring of GJ 436 revealed a 44.1 day rotational period and a 7.4 year activity cycle. Both of these are consistent with previous characterization of the star that found low activity and a moderate age. This information is especially helpful in constraining the role of changing starspots and plages on transit spectra and for the combination of data at different epochs. However, the expected color variation with increasing activity (i.e., an increase in star spot coverage making the star redder as it gets dimmer) is not found, possibly suggesting an interplay between star spots and stellar plages as GJ 436's activity changes. Furthermore, JWST will observe secondary eclipses of these planets as part of its GTO programs. Understanding how the stellar flux is changing in time is critical for an accurate determination of the flux ratio between the planet and the star.

Finally, we advocate for future JWST transit observations of GJ 436b to distinguish between high metallicity and cloudy atmosphere scenarios and provide information on disequilibrium chemistry and internal heating. Additionally, transit observations will compliment planned secondary eclipse observations by constraining the conditions at the terminator rather than the average dayside atmosphere.
 
\acknowledgments
We thank the anonymous referee for useful comments and suggestions. We thank David Sing and Nikolay Nikolov for helpful discussion. We also thank Ron Gilliland for help with the observing proposal. This work is based on observations made with the NASA/ESA Hubble Space Telescope, obtained from the data archive at the Space Telescope Science Institute. STScI is operated by the Association of Universities for Research in Astronomy, Inc. under NASA contract NAS 5-26555. Support for this work was provided by NASA through grants HST-GO-13308 and HST-GO-13665 from the Space Telescope Science Institute, which is operated by AURA, Inc., under NASA contract NAS 5-26555. G.W.H. acknowledges support from NASA, NSF, Tennessee State University, and the State of Tennessee through its Centers of Excellence Program.  An allocation of computer time from the UA Research Computing High Performance Computing (HPC) and High Throughput Computing (HTC) at the University of Arizona is gratefully acknowledged. This research made use of Astropy, a community-developed core Python package for Astronomy \citep{astropy:2013}, and PyAstronomy.\footnote{https://github.com/sczesla/PyAstronomy} This research has made use of NASA's Astrophysics Data System. IRAF is distributed by the National Optical Astronomy Observatory, which is operated by the Association of Universities for Research in Astronomy (AURA) under a cooperative agreement with the National Science Foundation. 
 
\bibliographystyle{apj}

\end{document}